\documentclass[fleqn,usenatbib,useAMS]{mnras}

\usepackage{graphicx}	
\usepackage{amsmath}	
\usepackage{amssymb}	
\usepackage{multicol}        
\usepackage{bm}		
\usepackage{pdflscape}	
\usepackage{subfig}
\usepackage{enumerate}

\usepackage{etoolbox}
\makeatletter
\patchcmd\@combinedblfloats{\box\@outputbox}{\unvbox\@outputbox}{}{\errmessage{\noexpand patch failed}}
\makeatother

\newcommand{\kms}{\,km\,s$^{-1}$} 

\def\kmsmpc{km s$^{-1}$ Mpc$^{-1}$}

\def\ergscm{erg s$^{-1}$ cm$^{-2}$}
\def\msun{\ifmmode M_{\odot} \else $M_{\odot}$\fi}
\def\msunyr{\ifmmode M_{\odot}~{\rm yr}^{-1} \else M$_{\odot}$~yr$^{-1}$\fi}
\def\zsun{\ifmmode Z_{\odot} \else Z$_{\odot}$\fi}
\def\lsun{\ifmmode L_{\odot} \else L$_{\odot}$\fi}

\newcommand{\mstar}{\ifmmode M_\star \else $M_\star $\fi}
\newcommand{\luv}{\ifmmode L_{\rm UV} \else $L_{\rm UV}$\fi}
\newcommand{\lir}{\ifmmode L_{\rm IR} \else $L_{\rm IR}$\fi}
\newcommand{\lbol}{\ifmmode L_{\rm bol} \else $L_{\rm bol}$\fi}

\usepackage[T1]{fontenc}
\usepackage{ae,aecompl}

\usepackage{newtxtext,newtxmath}

\title[KLASS: Kinematics of galaxies at cosmic noon]{The KMOS Lens-Amplified Spectroscopic Survey (KLASS) : Kinematics and clumpiness of low-mass galaxies at cosmic noon}

\author[]{
M. Girard$^{1,2,3}$\thanks{Email: mgirard@swin.edu.au},
C. A. Mason$^{4}$\thanks{Hubble Fellow},
A. Fontana$^{5}$,
M. Dessauges-Zavadsky$^{2}$,
T. Morishita$^{6}$,\newauthor
R. Amor\'in$^{7,8}$,
D. B. Fisher$^{1,3}$,
T. Jones$^{9}$,
D. Schaerer$^{2,10}$,
K. B. Schmidt$^{11}$,
T. Treu$^{12}$,\newauthor
\& B. Vulcani$^{13}$
\\
$^{1}$ Centre for Astrophysics and Supercomputing, Swinburne University of Technology, P.O. Box 218, Hawthorn, VIC 3122, Australia\\
$^{2}$ Observatoire de Gen\`eve, Universit\'e de Gen\`eve, 51 Ch. des Maillettes, 1290 Sauverny, Switzerland\\
$^{3}$ ARC Centre of Excellence for All Sky Astrophysics in 3 Dimensions (ASTRO 3D)\\
$^{4}$ Center for Astrophysics \,|\, Harvard \& Smithsonian, 60 Garden St, Cambridge, MA, 02138, USA \\
$^{5}$ INAF Osservatorio Astronomico di Roma, Via Frascati 33, 00040 Monteporzio (RM), Italy\\
$^{6}$ Space Telescope Science Institute, 3700 San Martin Drive, Baltimore, MD 21218, USA\\
$^{7}$ Instituto de Investigaci\'on Multidisciplinar en Ciencia y Tecnolog\'ia, Universidad de La Serena, Ra\'ul Bitr\'an 1305, La Serena, Chile\\
$^{8}$Departamento de F\'isica y Astronom\'ia, Universidad de La Serena, Av. Juan Cisternas 1200 Norte, La Serena, Chile\\
$^{9}$ University of California Davis, 1 Shields Avenue, Davis, CA 95616, USA\\
$^{10}$ CNRS, IRAP, 14 Avenue E. Belin, 31400 Toulouse, France\\
$^{11}$Leibniz-Institut f\:ur Astrophysik Potsdam (AIP), An der Sternwarte 16, 14482, Potsdam, Germany\\
$^{12}$ Department of Physics and Astronomy, UCLA, 430 Portola Plaza, Los Angeles, CA 90095-1547, USA\\
$^{13}$ INAF - Osservatorio Astronomico di Padova, Vicolo Osservatorio 5, IT-35122, Padova, Italy
}

\date{Accepted XXX. Received YYY; in original form ZZZ}

\pubyear{2019}

\begin{document}
\label{firstpage}
\pagerange{\pageref{firstpage}--\pageref{lastpage}}
\maketitle

\begin{abstract}
We present results from the KMOS Lens-Amplified Spectroscopic Survey (KLASS), an ESO Very Large Telescope (VLT) large program using gravitational lensing to study the spatially resolved kinematics of 44 star-forming galaxies at $0.6<z<2.3$ with a stellar mass of 8.1<log(\mstar/\msun)<11.0. These galaxies are located behind six galaxy clusters selected from the \textit{HST} Grism Lens-Amplified Survey from Space (GLASS).
We find that the majority of the galaxies show a rotating disk, but most of the rotation-dominated galaxies only have a low $\upsilon_{rot}/\sigma_0$ ratio (median of the sample of $\upsilon_{rot}/\sigma_0\sim2.5$).
We explore the Tully-Fisher relation by adopting the circular velocity, $V_{circ}=(\upsilon_{rot}^2+3.4\sigma_0^2)^{1/2}$, to account for pressure support. We find that our sample follows a Tully-Fisher relation with a positive zero-point offset of +0.18 dex compared to the local  relation, consistent with more gas-rich galaxies that still have to convert most of their gas into stars. 
We find a strong correlation between the velocity dispersion and stellar mass in the KLASS sample. When combining our data to other surveys from the literature, we also see an increase of the velocity dispersion with stellar mass at all redshift. We obtain an increase of $\upsilon_{rot}/\sigma_0$ with stellar mass at $0.5<z<1.0$. This could indicate that massive galaxies settle into regular rotating disks before the low-mass galaxies. For higher redshift ($z>1$), we find a weak increase or flat trend.
We investigate the relation between the rest-frame UV clumpiness of galaxies and their global kinematic properties. We find no clear trend between the clumpiness and the velocity dispersion and $\upsilon_{rot}/\sigma_0$. This could suggest that the kinematic properties of galaxies evolve after the clumps formed in the galaxy disk or that the clumps can form in different physical conditions.
\end{abstract}

\begin{keywords}
galaxies: evolution -- galaxies: kinematics and dynamics -- galaxies: high-redshift
\end{keywords}



\section{Introduction}

Galaxies at redshift $z\sim2$, at the peak of the cosmic star formation rate (SFR) density \citep{Madau2014}, show important differences compared to the galaxies in the local Universe. 
Deep \textit{HST} imaging has revealed that these distant galaxies have a complex morphology with star-forming clumps in their disks \citep[e.g.,][]{Elmegreen2005, Elmegreen2007, Livermore2012}. It has been recently found that the clumpy galaxy fraction peaks at a similar epoch ($1<z<3$), with more than $\sim50\%$ of the galaxies showing clumps in their internal structure \citep{Shibuya2016} and that the star formation rate of these clumps can be $\sim100-500$ times higher than in local HII regions \citep[e.g.][]{Genzel2011,Guo2012,Livermore2015}. 

Using infrared integral field spectroscopy with instruments such as SINFONI/VLT, KMOS/VLT and OSIRIS/Keck, several studies have analysed the kinematics of galaxies \citep[e.g.,][]{ForsterSchreiber2009,Epinat2012,Wisnioski2015,Stott2016}, revealing that most of the clumpy galaxies actually show ordered rotation. This is in contradiction with the initial idea that clumps were associated with interactions or mergers and suggests that they are formed by internal processes, such as the fragmentation of the galaxy disk due to gravitational instabilities \citep{Dekel2009}. Another key result from the large surveys is the high velocity dispersion ($\sigma_0$) observed in these distant systems, indicating that turbulent, gas-rich galaxies are present at this epoch. This suggests a large fraction of pressure supported galaxies at $z\sim1-3$ compared to the local Universe \citep[e.g.,][]{Epinat2010, Swinbank2017,Turner2017, Johnson2018, Ubler2019}. One possibility is that the high gas fraction of galaxies causes turbulence due to gravitational instabilities \citep[e.g.,][]{Genzel2011,Bournaud2016, Stott2016}. It has also been suggested that feedback from star formation plays a major role \citep[e.g.,][]{Green2010}. However, it remains unclear how the clumps form and evolve over time in these rotating disks and what really drives star formation and turbulence at this epoch.

The relation between the rotation velocity ($\upsilon_{rot}$) and stellar mass (\mstar) of galaxies, the Tully-Fisher relation \citep{Tully1977}, has also been studied at higher redshift to establish if a cosmic evolution of this relation exists \citep[e.g.,][]{Kassin2007,Harrison2017, Ubler2017,Straatman2017, Simons2017, Turner2018, Tiley2019}. An evolution would indicate a change in the dynamical to stellar mass ratio, giving information on when and how the gas mass was converted into stars, but also on the variation of the baryonic to dark matter fraction. 
However, the existence of this evolution is still debated since different results have been reported in the literature with sometime a negative, positive or even no offset of the zero-point of the relation. 

These previous large kinematic surveys in the infrared have provided important insight into the evolution and physical conditions of the galaxies around $z\sim1-3$, but they have focused only on relatively massive galaxies with log(\mstar/\msun)>10. Therefore, the evolution of the low-mass galaxies with log(\mstar/\msun)<10 is still not well established. It has been recently possible to probe the kinematics at lower stellar masses using deep observations \citep[e.g.,][]{Gnerucci2011,Kassin2012,Contini2016,Simons2017, Swinbank2017, Turner2017} and with the help of gravitational lensing \citep[e.g.,][]{Livermore2015, Leethochawalit2016, Mason2017, Girard2018a, Fontana2019}. 
Most of these low-mass galaxy surveys found lower rotation-dominated fraction (<50\%) and velocity and velocity dispersion maps that are more disturbed than in surveys focusing on more massive galaxies.  
Strong gravitational lensing also allows to probe the kinematics and clumpy morphology at high spatial resolution ($\lesssim1$ kpc) \citep[e.g.,][]{Jones2010, Livermore2015, Leethochawalit2016, Girard2018b, Patricio2018, Girard2019}. Such an upclose view of high-redshift galaxies can provide key information on the internal processes driving the star formation and clump formation. 

In this work, we present results from the full sample of the KMOS Lens-Amplified Spectroscopic Survey (KLASS), following a first paper presenting a sub-sample of these galaxies \citep{Mason2017}. The full sample allows us to investigate the evolution of the kinematic properties with stellar mass and UV clumpiness. We use gravitational lensing to probe the kinematics of typical galaxies at $0.6<z<2.3$ with stellar mass of 8.1<log(\mstar/\msun)<11. We are able to resolve the kinematics of 44/52 galaxies for which the magnification ($\mu \sim 2$) allows us to get a better spatial resolution than typical seeing-limited KMOS surveys and study lower mass galaxies (median of log(\mstar/\msun)=9.5).

The paper is organized as follows. Section \ref{sec:KLASS} describes the survey and target selection and Sect. \ref{sec:observations} the observations and data reduction. In Sect. \ref{sec:measurements}, we present the measurements of the integrated properties following in Sect. \ref{sec:kinematics} by the kinematic modeling and classification. Our analysis and main results on the Tully-Fisher and evolution of the kinematic properties with stellar mass and UV clumpiness are discussed in Sect. \ref{sec:analysis}.  We finally present our conclusions in Sect. \ref{sec:conclusion}.

In this paper, we use a cosmology with $H_0 = 70 $\kmsmpc, $\Omega_M = 0.3$, and $\Omega_\Lambda = 0.7$. When using values calculated with the initial mass function (IMF) of \citet{Salpeter1955}, we correct by a factor of 1.7 to convert to a \citet{Chabrier2003} IMF.

\section{The KMOS Lens-Amplified Spectroscopic Survey}
\label{sec:KLASS}

The KMOS Lens-Amplified Spectroscopic Survey (KLASS) is an ESO VLT KMOS Large Program \citep[196.A-0778, PI: A. Fontana,][]{Fontana2019} targeting the fields of six
massive galaxy clusters: 
Abell 2744;
MACS J0416.1-2403; 
MACS J1149.6+2223; 
MACS J2129.4-0741; 
RXC J1347.5-1145; 
RXC J2248.7-4431 (aka Abell S1063).
These fields have deep, multi-band \textit{HST}/ACS and WFC3, \textit{Spitzer}/IRAC and VLT/HAWKI imaging through the Cluster Lensing And Supernova survey with Hubble \citep[CLASH,][]{Postman2012}, Spitzer UltRa Faint SUrvey Program \citep[SURFSUP,][]{Bradac2014,Ryan2014,Huang2016} and Hubble Frontier Fields program \citep{Lotz2017}. KLASS observations were carried out in Service Mode during Periods 96-99 (October 2015 - October 2017).

The key science drivers of KLASS are:
\begin{enumerate}
    \item To probe the internal kinematics of galaxies at $z\sim1-3$, gravitationally lensed by the clusters, providing superior spatial resolution compared to surveys in blank fields \citep{Mason2017}.
    \item To measure $z > 7$ Lyman-$\alpha$ emission in Lyman-break galaxies, providing constraints on the timeline of cosmic reionization \citep{Mason2018,Mason2019}.
\end{enumerate}

This paper addresses the first science driver by presenting spatially resolved kinematics in the full KLASS sample. \citet{Mason2017} presented kinematics of 25 galaxies from 4 of the 6 KLASS fields with $1 - 5$ hour integrations. Here we present kinematics in 44 galaxies across all 6 fields, using the full $\sim10$ hour integrations.

\subsection{Target selection}
\label{sec:KLASS_targets}

KLASS targets were selected from the Grism Lens-Amplified Survey from Space\footnote{\url{http://glass.astro.ucla.edu}} \citep[GLASS,][]{Schmidt2014,Treu2015}. GLASS was a large Hubble Space Telescope (\textit{HST}) program which obtained grism spectroscopy of the fields of ten massive galaxy clusters. GLASS used the Wide Field Camera 3 (WFC3) G102 and G141 grisms to obtain spectra over the wavelength range $0.8-1.6$\,$\mu$m, with spectral resolution $R\sim150$. The GLASS data are publicly available, including \textit{HST} grism spectra for all the galaxies presented in this paper\footnote{\url{https://archive.stsci.edu/prepds/glass/}}.

KLASS $z\sim1-3$ targets were selected from the GLASS sample based on having at least one bright nebular emission line (H$\alpha$, [OIII] or [OII]) falling within the KMOS YJ range ($1-1.35$\,$\mu$m), and away from bright OH sky lines.

\section{Observations and data reduction}
\label{sec:observations}

A full description of the KLASS observations and data reduction procedures are given by \citet{Mason2019}, we provide a brief overview here. 

KLASS uses the KMOS YJ band ($1-1.35$\,$\mu$m), with spectral resolution $R\sim3400$. Each 24 integral field unit (IFU) of KMOS has 14$\times$14 spaxels of $0.2''\times0.2''$. Observations were carried out in service mode during Periods 96-99 (October 2015 - October 2017). Observations were executed in one hour observing blocks, where each block consisted of 1800s of science integration and 900s on sky, using alternating ABA science-sky-science integrating units. The final exposure time was $\sim10$ h per target.

Science frames were dithered by one pixel ($0.2''$) between frames. One star was observed in every field to monitor the point spread function (PSF) and the accuracy of dither offsets (except A2744, due to a broken IFU, a bright galaxy target was used to monitor dither offsets in this case). The data were reduced with the ESO KMOS pipeline v.1.4.3 \citep{Davies2013} with additional methods to improve sky subtraction for faint emission line sources \citep[see][]{Mason2019}. We combined exposures where the seeing was $\leq 0.8''$, the median seeing of our observations was $\sim0.6''$. This corresponds to a spatial resolution of $\sim5/\sqrt{\mu}$ proper kpc at $z\sim1$ (where $\mu$ is the gravitational lensing magnification of the galaxy). Frames were combined via sigma clipping, using the positions of the observed stars to determine the spatial position for combination.

\section{Measurements}
\label{sec:measurements}

\begin{table*}
\caption{Galaxy properties}             
\label{table_properties}      
\centering          
\begin{tabular}{l c c c c c c }   
\hline\hline       
Objects & Coordinates J2000 &  $z_{\rm spec}$ & Emission & $\mu$ &  log(M$_\star$) & SFR$_{\rm SED}$
\\
& RA - Dec & & lines & & [M$_{\odot}$] & [M$_{\odot} \, \mathrm{yr}^{-1}$] 
\\ 
\hline   
A2744-928 & $3.600359  \,  -30.396143 $ & 0.95 &  H$\alpha$  & $1.81_{-0.07}^{+0.19}$ & $ 9.83_{-0.02}^{+0.15}$ & $16.6_{-3.3}^{+0.6} $\\
A2744-892 & $3.599938  \,  -30.393555 $ & 1.34 &  [OIII], H$\beta$  & $1.96_{-0.29}^{+1.05}$ & $ 9.22_{-0.09}^{+0.10}$ & $7.9_{-1.1}^{+1.1} $\\
A2744-188 & $3.573398  \,  -30.380770  $ & 1.76 &  [OII], H$\beta$  & $2.41_{-0.62}^{+0.70}$ & $ 10.18_{-0.01}^{+0.09}$ & $58.6_{-2.7}^{+2.2} $\\
A2744-642 & $3.578407  \,  -30.388847 $ & 1.16 &  [OIII],H$\beta$  & $2.97_{-0.79}^{+1.04}$ & $ 9.80_{-0.16}^{+0.01}$ & $36.8_{-1.3}^{+5.0} $\\
A2744-692 & $ 3.575912 \,  -30.390070  $ & 1.16 &  [OIII], H$\beta$  & $2.63_{-0.55}^{+0.35}$ & $ 8.95_{-0.16}^{+0.03}$ & $4.3_{-0.6}^{+1.0} $\\
A2744-1773 & $ 3.57690 \,  -30.410185 $ & 1.65 &  [OIII]  &         $1.82_{-0.02}^{+0.35}$ & $ 9.46_{-0.03}^{+0.11}$ & $7.1_{-1.0}^{+0.8} $\\
A2744-1991 & $3.599785 \,  -30.413939 $ & 1.77 &  [OII], H$\beta$  & $3.03_{-0.38}^{+0.95}$ & $ 9.75_{-0.08}^{+0.01}$ & $20.8_{-0.8}^{+1.2} $\\
\hline  
MACS0416-268 & $64.033089   \,   -24.056322  $ & 1.37 &  [OIII], H$\beta$  & $1.89_{-0.35}^{+0.20}$  & $9.77_{-0.08}^{+0.02}$  & $18.3_{-0.9}^{+1.1} $ \\
MACS0416-1215 & $64.053574   \,   -24.065977 $ & 2.10 &  [OII]  & $2.88_{-0.96}^{+0.25}$   & $9.33_{-0.08}^{+0.10}$  & $6.6_{-0.5}^{+0.7} $\\
MACS0416-1174 & $64.041855   \,   -24.075808  $ & 1.99 &  [OII], H$\delta$, H$\gamma$  & $3.21_{-0.35}^{+1.04}$   & $9.39_{-0.09}^{+0.01}$  & $1.96_{-0.1}^{+0.1} $\\
MACS0416-1197 & $64.036591   \,   -24.067278  $ & 0.94 &  H$\alpha$, [NII], [SII]  & $17.89_{-2.70}^{+10.05}$   & $9.11_{-0.09}^{+0.02}$  & $2.0_{-0.1}^{+0.1} $\\
MACS0416-1321 & $64.016907   \,   -24.074207 $ & 1.63 &  [OIII]  & $2.98_{-1.10}^{+0.22}$   & $9.46_{-0.10}^{+0.02}$  & $4.5_{-0.3}^{+0.3} $\\
MACS0416-394 & $64.031036   \,   -24.078957 $ & 1.63 &  [OIII], H$\beta$   & $2.22_{-0.72}^{+0.35}$   & $9.93_{-0.01}^{+0.08}$  & $50.5_{-2.4}^{+0.1} $\\
MACS0416-1011 & $64.035187  \,    -24.070971 $ & 1.99 &  [OII], H$\delta$, H$\gamma$  & $2.37_{-0.74}^{+0.57}$   & $9.88_{-0.20}^{+0.01}$  & $3.1_{-0.1}^{+0.3} $\\
MACS0416-693 & $64.051437   \,   -24.071272$ & 1.35 &  [OIII], H$\beta$  & $2.11_{-0.63}^{+0.20}$   & $8.51_{-0.09}^{+0.02}$  & $0.62_{-0.14}^{+0.18} $\\
MACS0416-248 & $64.038254   \,   -24.056345$ & 0.85 &  H$\alpha$   & $1.63_{-0.15}^{+0.10}$   & $9.00_{-0.01}^{+0.02}$  & $1.5_{-0.2}^{+0.1} $\\
\hline  
MACS1149-593 & $177.406921  \,  +  22.407505$ & 1.48 &  H$\gamma$, [OIII], H$\beta$   & $1.43_{-0.03}^{+0.90}$   & $9.44_{-0.01}^{+0.07}$  & $60.8_{-3.4}^{+0.1} $\\
MACS1149-683 & $177.397232   \,  +   22.406185$ & 1.68 &  [OIII], H$\beta$  & $2.84_{-0.04}^{+0.90}$   & $8.15_{-0.01}^{+0.09}$  & $0.67_{-0.15}^{+0.12} $\\
MACS1149-691 & $177.382355   \,   +  22.405790$ & 0.98 &   H$\alpha$, [SII]  & $1.42_{-0.13}^{+0.38}$   & $9.56_{-0.01}^{+0.02}$  & $5.7_{-0.4}^{+0.1} $\\
MACS1149-1625 & $177.389999  \,    +  22.389452$ & 0.96 &   H$\alpha$, [SII]  & $1.26_{-0.10}^{+0.35}$   & $10.33_{-0.07}^{+0.02}$  & $102.1_{-3.5}^{+4.8} $\\
MACS1149-1237 & $ 177.384598   \,  +   22.396738$ & 0.70 &  H$\alpha$  & $1.13_{-0.03}^{+0.15}$  & $8.69_{-0.01}^{+0.01}$  & $1.2_{-0.1}^{+0.1} $\\
MACS1149-1644 & $177.394424   \,  +   22.389170$ & 0.96 &  H$\alpha$, [NII], [SII]   & $1.31_{-0.03}^{+0.35}$  & $10.28_{-0.02}^{+0.01}$  & $45.9_{-1.8}^{+1.8} $\\
MACS1149-1931 & $177.403442   \,  +   22.381588$ & 1.41 &  [OIII], H$\beta$   & $1.53_{-0.09}^{+0.15}$  & $10.46_{-0.14}^{+0.01}$  & $216.2_{-7.1}^{+18.1} $\\
MACS1149-1757 & $177.408539   \,  +   22.386808$ & 1.25 &  [OIII], H$\beta$  & $2.17_{-0.03}^{+0.60}$  & $8.26_{-0.01}^{+0.02}$  & $0.55_{-0.14}^{+0.16} $\\
MACS1149-1501 & $177.397034   \,  +   22.396011$ & 1.49 &  [OIII], H$\beta$   & $5.42_{-0.13}^{+1.03}$  & $9.90_{-0.08}^{+0.01}$  & $56.9_{-0.1}^{+2.5} $\\
\hline  
MACS2129-1705 & $322.355498   \,  -7.707086$ & 1.05 &  H$\alpha$  & $1.3_{-0.03}^{+0.04}$  & $10.89_{-0.09}^{+0.05}$  & $96.7_{-6.2}^{+6.2} $\\
MACS2129-994 & $322.348023   \,  -7.692095$ & 0.60 &  H$\alpha$, [NII], [SII]   & $1.02_{-0.03}^{+0.03}$  & $9.65_{-0.01}^{+0.01}$  & $5.6_{-0.1}^{+0.1} $\\
MACS2129-1847 & $322.352775  \,   -7.710149$ & 1.88 &   [OII], H$\delta$, H$\gamma$  & $1.38_{-0.04}^{+0.03}$  & $9.81_{-0.14}^{+0.08}$  & $74.1_{-4.0}^{+7.9} $\\
MACS2129-1833 & $322.362724  \,   -7.710150$ & 2.29 &  [OII] & $1.56_{-0.06}^{+0.07}$  & $9.96_{-0.14}^{+0.08}$  & $127.8_{-5.6}^{+7.6} $\\
MACS2129-1539 & $322.363350  \,   -7.703212$ & 1.65 &  [OIII], H$\beta$  & $1.74_{-0.05}^{+0.06}$  & $9.92_{-0.07}^{+0.01}$  & $48.9_{-1.8}^{+2.7} $\\
MACS2129-974 & $322.368223   \,  -7.692491$ & 1.54 &  [OIII], H$\beta$  & $4.37_{-0.89}^{+1.97}$  & $8.91_{-0.07}^{+0.15}$  & $17.7_{-2.8}^{+0.3} $\\
MACS2129-1126 & $322.358615   \,  -7.694884$ & 1.36 & [OIII] & $4.71_{-0.09}^{+0.12}$  & $11.05_{-0.02}^{+0.02}$  & $336.1_{-7.8}^{+0.1} $\\
MACS2129-877 & $322.380612   \,  -7.689563$ & 0.58 &  H$\alpha$, [SII]   & $1.0_{-0.03}^{+0.03}$  & $9.40_{-0.01}^{+0.01}$  & $3.1_{-0.1}^{+0.1} $\\
MACS2129-564 & $322.364817  \,   -7.683663$ & 1.37 &  [OIII], H$\beta$   & $1.83_{-0.03}^{+0.04}$  & $9.73_{-0.02}^{+0.04}$  & $16.5_{-1.5}^{+1.5} $\\
\hline  
RXJ1347-472 & $206.891012  \,  -11.747415$ & 0.91 &  H$\alpha$, [NII], [SII]   & $2.86_{-0.03}^{+0.04}$  & $9.29_{-0.01}^{+0.01}$  & $9.2_{-0.4}^{+0.0} $\\
RXJ1347-1230 & $206.871917  \,  -11.760966$ & 1.77 &  [OII] & $100.56_{-46.17}^{+226.24}$  & $8.52_{-0.07}^{+0.01}$  & $1.9_{-0.1}^{+0.1} $\\
RXJ1347-1419 & $206.875708  \,  -11.764457$ & 1.14 &  [OIII], H$\beta$   & $8.31_{-0.17}^{+0.19}$  & $9.44_{-0.08}^{+0.01}$  & $16.2_{-0.1}^{+0.8} $\\
RXJ1347-1261 & $206.888155  \,  -11.761284$ & 0.61 &  H$\alpha$, [NII]  & $1.62_{-0.03}^{+0.03}$  & $10.59_{-0.02}^{+0.03}$  & $30.9_{-1.1}^{+1.3} $\\
RXJ1347-795 & $206.896042 \,  -11.753651$ & 0.62 &  H$\alpha$, [NII], [SII]  & $1.48_{-0.03}^{+0.03}$  & $10.35_{-0.09}^{+0.01}$  & $28.1_{-0.1}^{+1.5} $\\
RXJ1347-450 & $ 206.900187  \,  -11.747581$ & 0.85 &  H$\alpha$, [NII], [SII]   & $1.88_{-0.04}^{+0.03}$  & $10.01_{-0.10}^{+0.01}$  & $24.3_{-0.1}^{+2.4} $\\
RXJ1347-287 & $206.902241   \, -11.744268$ & 1.00 &  H$\alpha$, [NII]  & $2.31_{-0.06}^{+0.07}$  & $9.21_{-0.08}^{+0.02}$  & $6.1_{-0.4}^{+0.5} $\\
\hline  
RXJ2248-1006 & $342.174383  \,   -44.532903$ & 0.61 &  H$\alpha$, [NII], [SII]   & $5.35_{-1.15}^{+0.50}$  & $9.29_{-0.08}^{+0.01}$  & $7.4_{-0.1}^{+0.4} $\\
RXJ2248-1652 & $342.205097  \,   -44.544419 $ & 1.64 &  [OIII], H$\beta$    & $1.36_{-0.26}^{+0.09}$  & $9.60_{-0.01}^{+0.01}$  & $9.5_{-0.6}^{+0.6} $\\
RXJ2248-918 & $342.192466   \,  -44.530468$ & 1.26 &  [OIII] & $55.79_{-20.25}^{+20.16}$  & $8.65_{-0.13}^{+0.03}$  & $0.23_{-0.01}^{+0.03} $\\
RXJ2248-943 & $342.195250   \,  -44.527900$ & 1.23 &  [OIII], H$\beta$   & $44.60_{-24.05}^{+1.75}$  & $7.99_{-0.08}^{+0.02}$  & $0.19_{-0.01}^{+0.02} $\\
RXJ2248-662 & $342.187756    \,  -44.527445$ & 1.40 &  [OIII], H$\beta$   & $1.02_{-0.03}^{+0.21}$  & $9.91_{-0.02}^{+0.09}$  & $24.4_{-2.8}^{+2.9} $\\
RXJ2248-428 & $342.186462   \,  -44.521186 $ & 1.23 &  [OIII], H$\beta$    & $2.98_{-0.04}^{+0.23}$  & $9.40_{-0.01}^{+0.01}$  & $3.1_{-0.2}^{+0.1} $\\
RXJ2248-159 & $342.174107    \, -44.514794$ & 0.97 &  H$\alpha$, [NII], [SII] & $1.40_{-0.21}^{+0.05}$  & $9.43_{-0.02}^{+0.02}$  & $6.5_{-0.4}^{+0.4} $\\
RXJ2248-332 & $342.167043    \, -44.518778$ & 1.43 &  [OIII], H$\beta$    & $1.63_{-0.49}^{+0.07}$  & $8.52_{-0.01}^{+0.07}$  & $0.64_{-0.08}^{+0.01} $\\
\noalign{\smallskip}
\hline                             
\end{tabular}
\end{table*}

\begin{figure}
\includegraphics[width=\columnwidth]{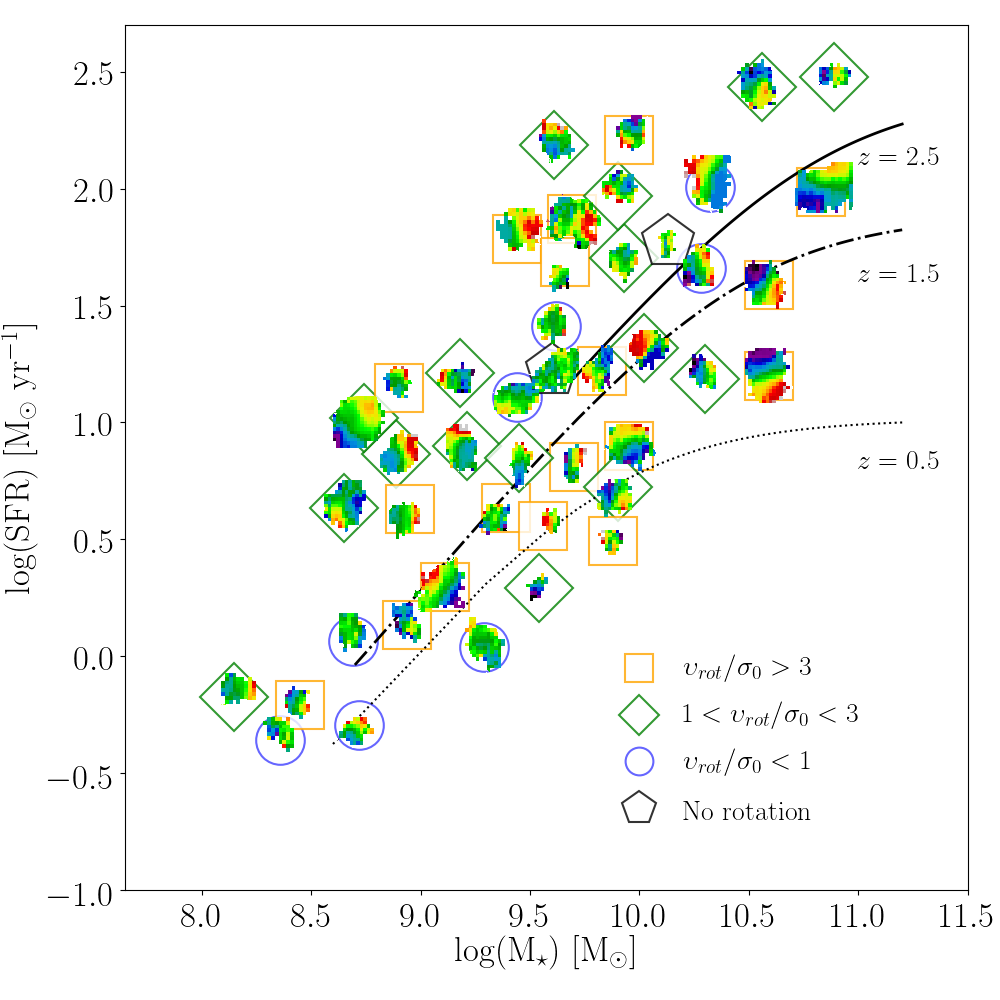}
%
\caption[]{Star formation rate as a function of the stellar mass. The stellar mass and SFR have been derived from SED fitting. The values are all lensing-corrected. The velocity maps of each galaxy with resolved kinematics is shown on the plot. The shape and color around the velocity maps represent the galaxy with no rotation (grey pentagon),  $\upsilon_{rot}/\sigma_0 < 1$ (blue circle),  $1<\upsilon_{rot}/\sigma_0 < 3$ (green  diamond), and $\upsilon_{rot}/\sigma_0>3$ (orange square). The curves represent the relation of SFR as a function of stellar mass derived by \citet{Tomczak2016} for the redshifts of 0.5, 1.5, and 2.5.}
\label{MS_klass}
\end{figure}

We detect emission lines for 49 galaxies among the total of 52 observed with KLASS. The main emission lines detected are the [OII], [OIII] and H$\alpha$ lines, but we also identify in several galaxies some fainter lines such as  H$\beta$, H$\gamma$, H$\delta$, [NII], and [SII]. These lines are listed in Table \ref{table_properties}. For these galaxies,
we create an integrated spectrum by summing spaxels in an aperture to maximize the signal-to-noise ratio. To measure the flux, the line position and line width, we fit 1000 Gaussian curves using a Monte Carlo technique perturbing the flux of the emission lines. The continuum level is determined by the average of all the pixels in a window of 50~\AA \, around the center of the emission line. In this window, all the pixels affected by skylines are masked. When the window is populated by too many skylines, we extend it to 70 \AA. As a result, we obtain an accurate systemic redshift, the [OII], [OIII] or H$\alpha$ flux, the full width half maximum (FWHM) of the lines, the integrated velocity dispersion for each galaxy and their associated uncertainties. The spectroscopic redshifts are in good agreement with the GLASS redshifts and are listed in Table \ref{table_properties}.


The stellar masses and SFR of each galaxy are derived from SED fitting using available photometry from the Hubble Frontier Fields and CLASH catalogs \citep{Morishita2017, Postman2012}. The Fitting an Assessment of Synthetic Templates (FAST) code \citep{Kriek2009} is used to perform the fit with the spectroscopic redshift derived from the integrated spectrum. The \citet{Bruzual2003} stellar populations are adopted together with an exponentially declining star formation history, a \citet{Chabrier2003} IMF and the dust attenuation law from \citet{Calzetti2000}. The values obtained for the stellar mass and SFR are then corrected for magnification using the SWUnited cluster mass models \citep{Bradac2005, Bradac2009} and the multiple models of the Hubble Frontiers Fields that are publicly available\footnote{\url{https://archive.stsci.edu/prepds/frontier/lensmodels/}} \citep{Hoag2016, Huang2016, Hoag2019}. Table \ref{table_properties} presents the magnifications, stellar masses, and SFR both corrected for the magnification. 


We find spectroscopic redshifts between $0.6<z<2.25$ for these 49 galaxies, including two sources at $z>2$. From the SED fitting, we find a stellar mass range of 8.1<log(\mstar/\msun)<11.0, with a median of  log(\mstar /\msun)=9.5. The SFR from the SED fitting for this sample is 0.2< SFR [\msunyr])<336, with a median of $10$ \msunyr. Fig. \ref{MS_klass} shows the star formation rate as a function of stellar mass corrected for the magnification. Only the 44 galaxies with resolved kinematics are shown on this plot with their velocity maps and kinematic classification (see Sect. \ref{sec:kinematics}).


\section{Kinematics}
\label{sec:kinematics}

\subsection{Modeling}


\begin{table*}
\caption{Kinematic properties }        
\label{kin_properties}      
\centering          
\begin{tabular}{l c c c c c }     
\hline\hline       
Objects & $\upsilon_{rot} $ & $\sigma_0$ & Kinematic  & Clumpiness    
\\
 & [\kms] & [\kms] & Classification &   &  
\\ 
\hline   
A2744-928 &  $ 328 \pm  17 $   &   $  20 \pm   $  6  &  1 & 0.22  \\
A2744-892 &  $ 99 \pm 9  $   &   $ 80 \pm  2 $  &  4 &  Potential merger   \\
A2744-188 &   Non-rotator     &   Non-rotator  &  3   &   $ -  $ \\
A2744-642 &  $ -   $   &   $  -  $  &  5  &  $ -  $ \\ 
A2744-692 &  $  145 \pm 18  $   &   $ 27 \pm 3  $ &  1 & 0.28  \\
A2744-1773 &  $ 62 \pm 3  $   &   $ 23 \pm  4 $ &  2   & 0.06 \\
A2744-1991 &  Non-rotator    &   Non-rotator   &  4   & Potential merger \\
\hline  
MACS0416-268 &  $  31 \pm  30 $   &   $ 36 \pm  3 $  & 3  & 0.22  \\
MACS0416-1215 &  $ -  $   &   $ -  $  &  5  &  $ -  $ \\ 
MACS0416-1174 &  $  69\pm  10 $   &   $ 44 \pm  15 $ &  4  &  Potential merger \\
MACS0416-1197 &  $ 215 \pm  12 $   &   $ 7 \pm  4 $  & 1  & 0  \\
MACS0416-1321 &  $ 230 \pm 13  $   &   $ 61 \pm  3 $  & 1  & 0.79   \\
MACS0416-394 &  $ 76 \pm 11  $   &   $32  \pm  1 $  &  2  & 0 \\
MACS0416-1011 &  $ 249 \pm  90 $   &   $ 42 \pm  12 $  & 4 &  Potential merger   \\
MACS0416-693 &  $ 140 \pm  30 $   &   $ 29 \pm  2 $  &  1 & 0   \\
MACS0416-248 &  $ 65 \pm  21 $   &   $ 21 \pm  7 $  &  1  & 0  \\
\hline  
MACS1149-593 &  $ 127 \pm  5 $   &   $ 13  \pm  4 $ & 1  & 0.24  \\
MACS1149-683 &  $ 49 \pm  4 $   &   $ 26 \pm  1 $  &  2  & 0.28  \\
MACS1149-691 &  $ 188 \pm 11  $   &   $ 60 \pm  4  $  &  1 & 0  \\
MACS1149-1625 &  $ 99 \pm 13  $   &   $ 139 \pm 5  $  & 4  &  Potential merger  \\
MACS1149-1237 &  $ 9 \pm  7 $   &   $ 35 \pm  3 $  &   3  & 0 \\
MACS1149-1644 &  $ 157 \pm 8  $   &   $ 170 \pm  10 $  & 3 & 0  \\
MACS1149-1931 &  $ 123 \pm  8 $   &   $ 98 \pm  2 $  &  2  &0.27 \\
MACS1149-1757 &  $ 15 \pm  6 $   &   $ 18 \pm  3 $  & 3  & 1.0  \\
MACS1149-1501 &  $ 208 \pm  41 $   &   $ 32 \pm  11 $  & 1  & 0.29 \\
\hline  
MACS2129-1705 &  $ 239 \pm 7  $   &   $ 51 \pm 2  $  &  1 & 0.87  \\
MACS2129-994 &   $ 168 \pm  17 $   &   $ 33 \pm  2 $  & 1  & 0.05  \\
MACS2129-1847 &   $ 243 \pm  41 $   &   $ 98 \pm  5 $ & 2  & 0.04  \\
MACS2129-1833 &   $ 306 \pm 23  $   &   $ 55 \pm  12 $ &  1 & 0.44 \\ 
MACS2129-1539 &   $ 77 \pm  18 $   &   $ 74 \pm  3 $  &  2  &0.58 \\
MACS2129-974 &    $ 113 \pm 17  $   &   $ 37 \pm  4 $  &  1  & 0 \\ 
MACS2129-1126 &    $ 135 \pm 8  $   &   $  76 \pm  6 $  &  2  & 0  \\
MACS2129-877 &    $ 25 \pm  6 $   &   $ 28 \pm 3  $  &  3  &0.34  \\
MACS2129-564 &    $ 195 \pm  5 $   &   $ 84 \pm 3  $ & 4  &  Potential merger  \\
\hline  
RXJ1347-472 &   $ 88 \pm 5  $   &   $ 58 \pm  1 $  & 2  & 0.25   \\
RXJ1347-1230 &   $  -  $   &   $ - $  &  5  &  $ -  $ \\
RXJ1347-1419 &   $ 44 \pm  22 $   &   $ 52 \pm  1 $ & 3  &0.12  \\
RXJ1347-1261 &   $ 227 \pm  5 $   &   $ 42 \pm  4 $ & 1   & 0 \\
RXJ1347-795 &   $  184\pm  2 $   &   $ 58 \pm  1 $  &  1 &0.64   \\
RXJ1347-450 &   $ 131 \pm  12 $   &   $ 92 \pm  5 $ & 2  &0.94  \\
RXJ1347-287 &   $ 77 \pm 12  $   &   $ 38 \pm  1 $  &  2  &0.19  \\
\hline  
RXJ2248-1006 &   $ 114 \pm  3 $   &   $ 58 \pm  1 $ &  4   &  Potential merger \\
RXJ2248-1652 &   $ 106 \pm  6 $   &   $ 50 \pm  2 $ &  2  &0.53  \\
RXJ2248-918 &   $  -  $   &   $ -   $ & 5 &  $ -  $  \\ 
RXJ2248-943 &   $ -  $   &   $  -  $  & 5 &  $ -  $  \\
RXJ2248-662 &   $ 51 \pm 7  $   &   $ 14 \pm 2  $ &  1  & 0.63  \\
RXJ2248-428 &   $ 95 \pm 21  $   &   $ 25 \pm 3  $  &  1 & 0.42   \\
RXJ2248-159 &   $  63\pm 12  $   &   $  35\pm   3$  & 2  & 0.92  \\
RXJ2248-332 &   $  22\pm 3  $   &   $  26\pm  2 $  & 4  & Potential merger   \\

\noalign{\smallskip}
\hline                                   
\end{tabular}
\end{table*}

To obtain the flux, velocity, and velocity dispersion maps, we perform Gaussian  fits on the [OII], [OIII] or H$\alpha$ emission line in individual spaxels. We use the spectroscopic redshifts obtained from the integrated spectrum to determine the central wavelength for the velocity maps. We use a spectral window of 50 \AA \, to measure the continuum and mask all the skylines. We also determine the level of noise for each spaxel in the same spectral window. We first fit the emission line in one spaxel ($0.2''\times 0.2''$) and bin spaxels in $0.4''\times 0.4'$ and $0.6''\times 0.6''$ successively if the signal-to-noise ratio is lower than five. We reject the spaxels when the signal-to-noise ratio is still lower than five after the binning. We apply this method for all the resolved galaxies in the sample. We obtain kinematic maps for 44 galaxies (see Fig. \ref{map_galaxies}). The other galaxies are unresolved or it was impossible to obtain kinematic maps for the whole galaxies due to very high magnification. We also note that the kinematics is not significantly affected by the shear caused by the lensing in seeing-limited observations for galaxies with a low magnification ($\mu \lesssim 5$). We find a median differential amplification of only $\sim0.1$ in our galaxies.

To obtain the kinematic properties, we model the kinematics of the galaxies using  GalPaK$^{\rm3D}$  \footnote{\url{http://www.ascl.net/1501.014}}, a three-dimensional galaxy disk modeling code that directly fits the datacubes \citep{Bouche2015}. The model is convolved with the PSF and the line spread function (LSF) to derive kinematic and morphological properties. The beam smearing is therefore taken into account when modelling the kinematics. The PSF has been determined from a star observed in one of the IFUs during the observations and the LSF has been measured from the skylines. We use an exponential radial flux profile and adopt an arctangent function for the velocity profile \citep{Courteau1997}. The disk half-light radius, the flux, the inclination, the position angle, the turnover radius,  the maximum rotation velocity, and the intrinsic velocity dispersion are free to vary. We are able to fit a kinematic model for 42 galaxies. It was not possible to model the kinematics for galaxies that do not show any velocity gradient in their velocity field. 
We also exclude galaxies with a magnification higher than $\sim5$ for our analysis (Sect. \ref{sec:analysis}) to avoid galaxies that could have a significant effect from the lensing. The error on the rotation velocity of the remaining galaxies due to the lensing is about $10\%$.
The inclination-corrected rotation velocity, $\upsilon_{rot}$, and velocity dispersion, $\sigma_0$, obtained from GalPaK$^{\rm3D}$ are presented in Table \ref{kin_properties}. The velocity maps obtained from the kinematic models, rotation curves and dispersion profiles of our sample are presented in Fig. \ref{map_galaxies}. 

We find a signal-to-noise ratio higher than five at 1.5 $\times R_e$, where $R_e$ is the effective radius, on the kinematic major axis in 77\% of our galaxies. This corresponds to the radius around which the rotation curve reaches its maximum and starts to flatten \citep[e.g.][]{Genzel2017, Lang2017}.
We also note that three of our targets (MACS1149-593, MACS1149-683, and MACS2129-1833) have also been observed with Keck OSIRIS using adaptive optics (AO) \citep{Hirtenstein2019}. The AO data show similar results to those in Table \ref{kin_properties}. Thus, the kinematics are reasonably well recovered with seeing-limited KMOS data, based on direct comparison to measurements with $\sim10$x higher resolution.


\begin{figure}
\includegraphics[width=\columnwidth]{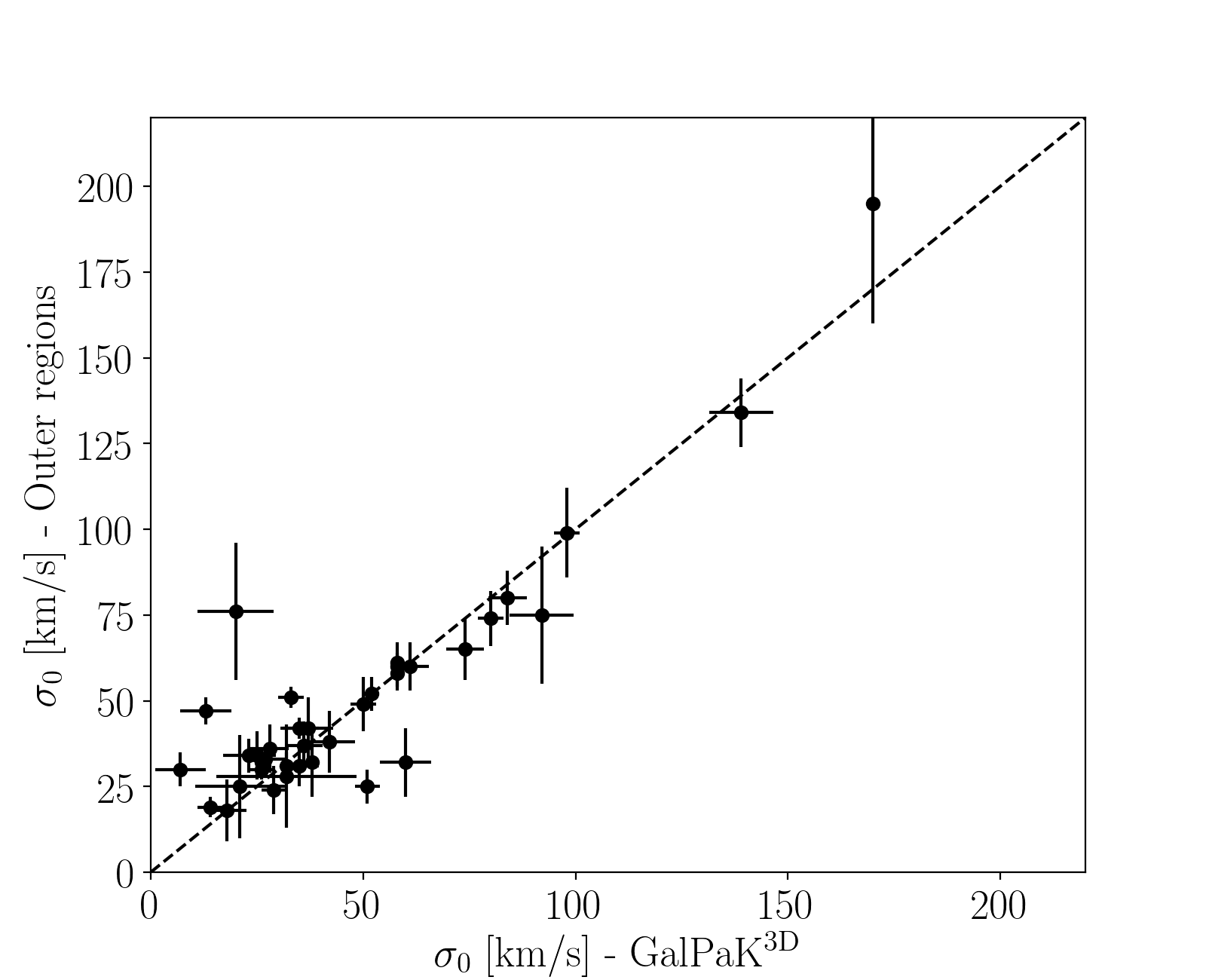}
%
%
\caption[]{Comparison of the instrinsic velocity dispersion measured from the outer regions and obtained with the three-dimensional modeling code GalPaK$\rm^{3D}$ \citep{Bouche2015}. The dashed line represents the 1:1 relation. }
\label{sigma_comparison_klass}
\end{figure}

\begin{figure*}
\subfloat{\includegraphics[width=\columnwidth]{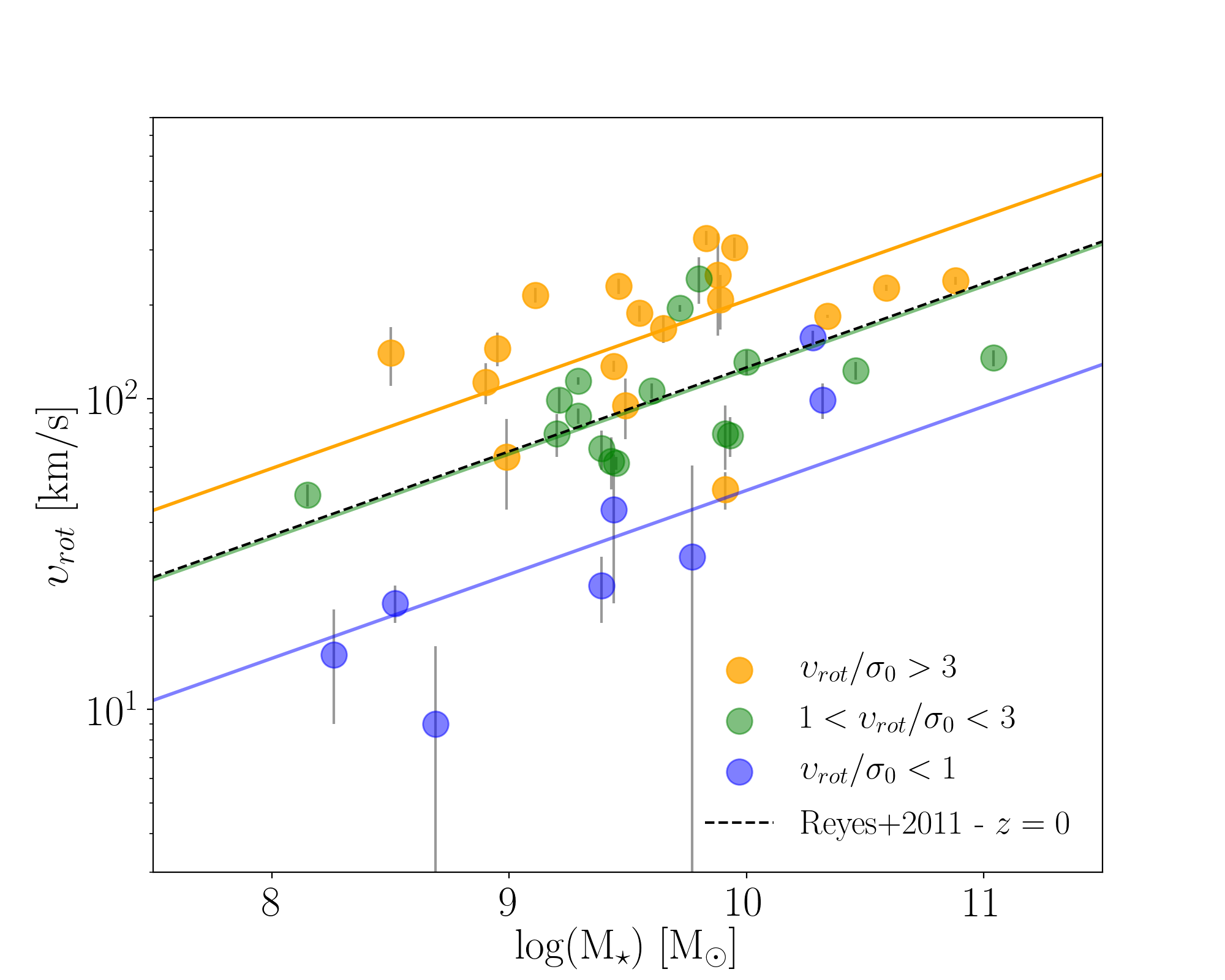}} 
\subfloat{\includegraphics[width=\columnwidth]{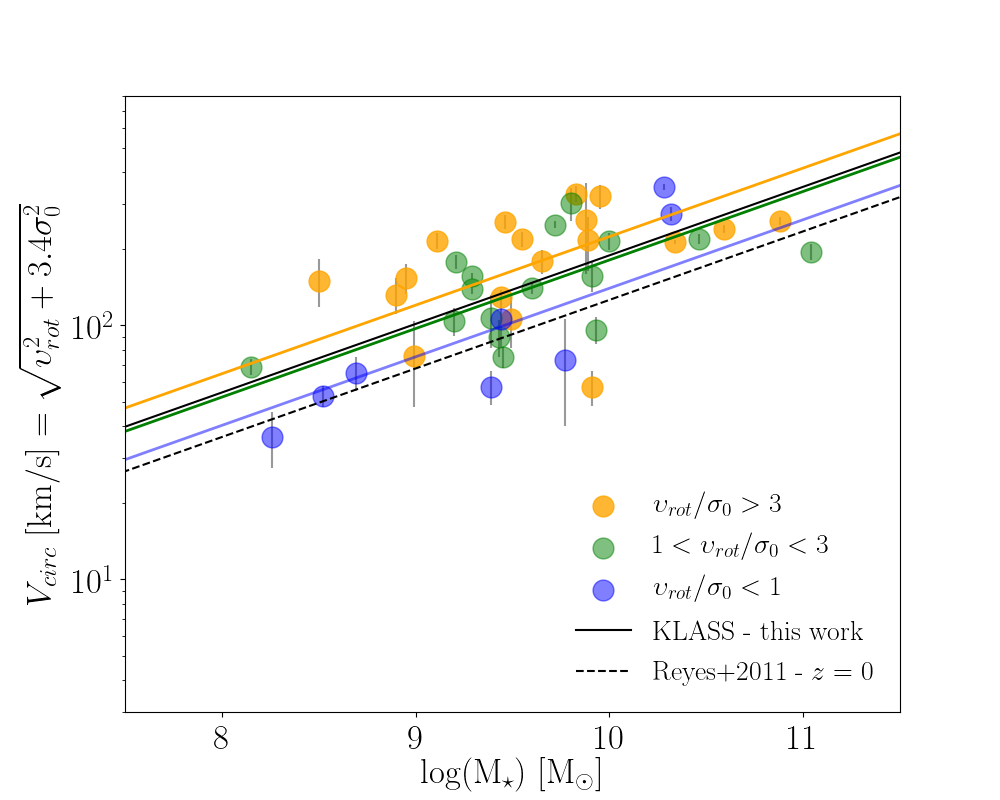}}
%
%
\caption[]{Rotation velocity (left panel) and circular velocity (right panel) as a function of stellar mass. The coloured lines represent the fit for each kinematic category. The black line presents the fit for the full sample, regardless of kinematic classification. The dashed black line indicates the Tully-Fisher relation observed in the local Universe by \citet{Reyes2011}. }
\label{TF_klass}
\end{figure*}

\subsection{Velocity dispersion}
\label{sec:veldispersion}

To verify the intrinsic velocity dispersion obtained with GalPaK$^{\rm3D}$, which assumes a spatially constant dispersion over the disk, we also perform a measurement directly from the velocity dispersion maps on the major axis in the outer regions of the galaxies, where the beam-smearing is known to be negligible \citep[e.g.,][]{ForsterSchreiber2009, Wisnioski2015}, especially for low-mass galaxies \citep[][]{Burkert2016,Johnson2018}. 
We then apply a correction in quadrature to account for the instrumental broadening.
We compare these measurements to the values obtained with GalPaK$\rm^{3D}$ in Fig. \ref{sigma_comparison_klass}. The values are consistent within the uncertainties for most of the galaxies. We find a large discrepancy (>20 \kms) for only five galaxies.
This could be due to asymmetry in the velocity maps observed in some galaxies, which could be caused by minor merger or interactions that are not clearly visible in the velocity map or \textit{HST} images, or to low intrinsic velocity dispersions, difficult to recover when the observed velocity dispersion is of the same order as the spectral resolution. 
For the rest of this work, we adopt the intrinsic velocity dispersion values from the three-dimensional modeling code GalPaK$\rm^{3D}$.

\subsection{Classification}

The kinematics of galaxies are generally classified using the ratio  $\upsilon_{rot}/\sigma_0$, with $\upsilon_{rot}/\sigma_0$> 1 for rotation-dominated galaxies and $\upsilon_{rot}/\sigma_0$<1 for dispersion-dominated galaxies \citep[e.g.][]{Wisnioski2015}. Using this definition, we find a rotation-dominated fraction of 77\% and a dispersion-dominated fraction of 18\%. We have also 2 galaxies, representing 5\% of the sample, that do not show any rotation in their velocity map. For our final classification, we divide the galaxies into the following kinematic categories:

\begin{enumerate}
\item[1.] Regular rotation: rotation-dominated galaxies with  $\upsilon_{rot}/\sigma_0$ > 3;
\item[2.] Irregular rotation: rotation-dominated galaxies with 1<$\upsilon_{rot}/\sigma_0$ < 3;
\item[3.] Dispersion-dominated :  pressure supported galaxies with $\upsilon_{rot}/\sigma_0$ < 1 and galaxies showing no velocity gradient in their velocity field;
\item[4.] Mergers: galaxies possibly showing sign of interactions;
\item[5.] Unresolved: We remove from the sample two unresolved galaxies and three galaxies for which no kinematic modeling has been performed due to the high magnification. These galaxies are not taken into account when deriving the rotation- and dispersion-dominated fractions. 
\end{enumerate}

Table \ref{kin_properties} shows the kinematic properties and classification of the 49 galaxies. We introduce here the irregular rotation category since many galaxies show a very low ratio with 1<$\upsilon_{rot}/\sigma_0$ <3. Indeed, the median of this ratio in our sample is $\upsilon_{rot}/\sigma_0\sim 2.5$. At $\upsilon_{rot}/\sigma_0\sim3$, the gas fraction needs to be higher than $50\%$ for disk to have a Toomre $Q$ lower than 1 \citep[marginal stability,][]{Toomre1964}. For thick disks, instability occurs at $Q\sim0.7$. The $Q\sim1$ value is for an infinitely thin disk \citep{Kim2007, Romeo2010}. From this classification, we find that $39\pm9$\% of the galaxies show a regular rotation, $27\pm7$\% are irregular rotators, and $16\pm6$\% are dominated by dispersion. 
If we extend our classification by looking at the \textit{HST} images, we are also able to identify which galaxies possibly show sign of a merging system. These galaxies are marked in Table \ref{kin_properties} and represent the remaining $18\pm7 \%$ of the sample. 

Similarly, \citet{Mason2017} found a rotation-dominated fraction of $\sim65\%$ with most of the galaxies being classified as irregular rotators. \citet{Girard2018a} found a lower rotation-dominated fraction of $\sim25-40\%$ for their sample of 24 low-mass galaxies at $1.4<z<3.5$, which is also comparable to the rotation-dominated fraction of $\sim34\%$ obtained by \citet{Turner2017} for their 32 galaxies at $z\sim3.5$. This could be due to the higher redshift of the galaxies studied in these two surveys compared to this work. \citet{Simons2017} obtained a decrease of the rotation-dominated fraction with redshift and with stellar mass, with a fraction of $\sim45-75\%$ expected for galaxies of 9.0<log(\mstar/\msun)<10.0 at $0.6<z<2.25$.


\section{Analysis and Discussion}
\label{sec:analysis}
In the following sections, we investigate the Tully-Fisher relation for these low-mass galaxies and the evolution of their kinematic properties with stellar mass. We also explore correlations between the kinematic properties and  morphology, more particularly the rest-frame UV clumpiness.

\subsection{Tully-Fisher relation}

The relation between the rotation velocity, which traces the dynamical mass of disk galaxies, and the stellar mass, also called the Tully-Fisher relation \citep[][]{Tully1977}, has been analyzed by many studies at high redshift \citep[e.g.,][]{Kassin2007, Simons2015, Tiley2016a, Harrison2017, Ubler2017, Turner2018}. However, the existence of an evolution of this relation with redshift is still debated. It also remains unclear if the Tully-Fisher relation depends on stellar mass since very few surveys studied low-mass galaxies, especially at $z>1$. For example, \citet{Harrison2017} determined that there is no evolution with redshift and stellar mass for their galaxies at $0.6<z<1.0$ with a stellar mass of $9.0<$log$(\mstar/\msun)<11.0$. \citet{Simons2015} found that the galaxies with log$(\mstar/\msun)<9.5$ do not follow the Tully-Fisher relation since they show a large scatter at $0.1 < z < 0.375$. In addition, \citet{Christensen2017} reported a power-law slope and normalisation independent of the redshift, but also found a break with a steeper slope for low-mass galaxies in their sample at $0<z<3$ with stellar mass of 7.0<log$(\mstar/\msun)$<11.5. In this section, we use KLASS to investigate the redshift and stellar mass evolution of the Tully-Fisher relation to extend these previous studies to low-mass galaxies at $0.6<z<2.3$.

\begin{figure*}
\includegraphics[scale=0.9]{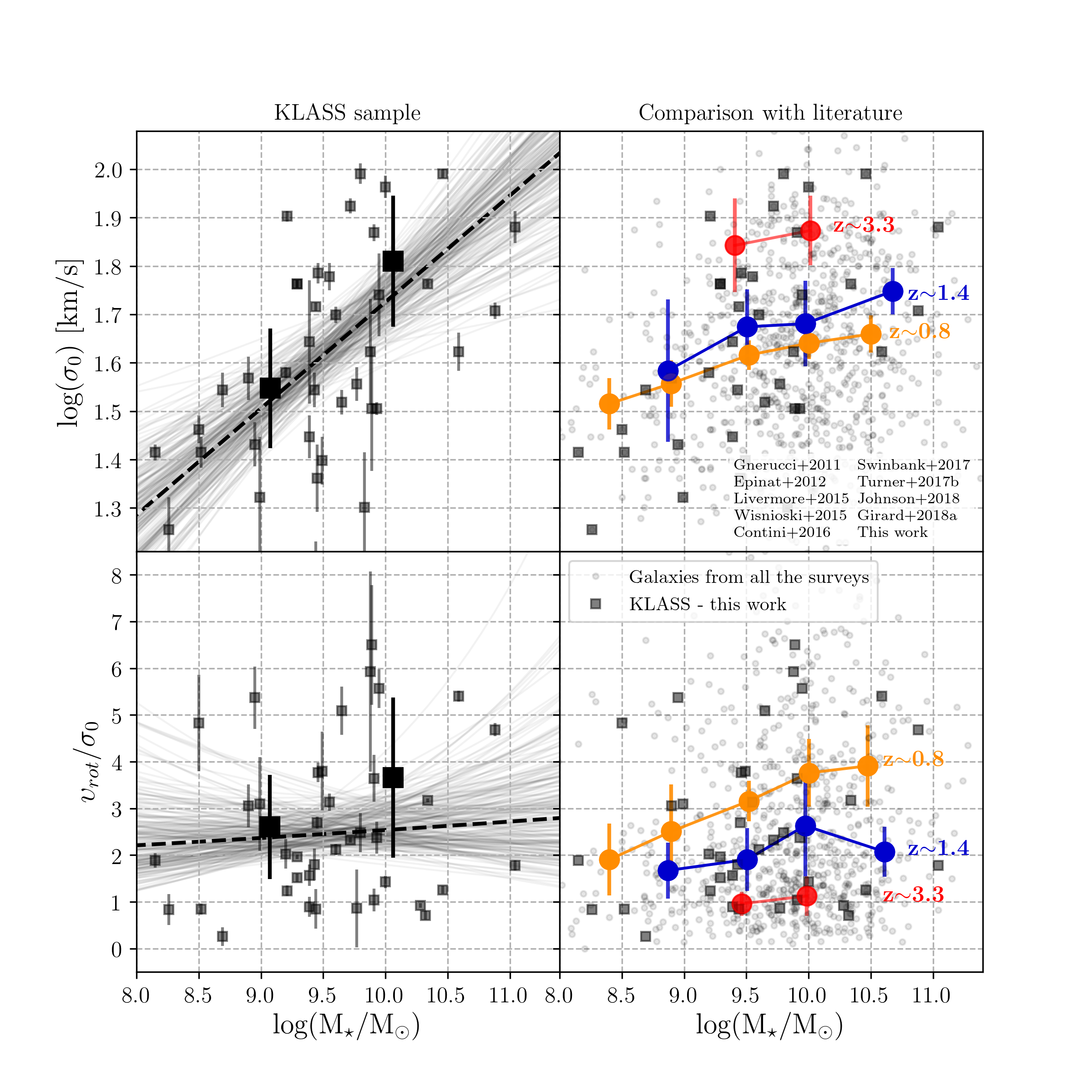}
\medskip
%
\caption[]{Intrinsic velocity dispersion and rotation velocity to intrinsic velocity dispersion ratio, $\upsilon_{rot}/\sigma_0$, as a function of stellar mass for KLASS (square) and a compilation of galaxies from different surveys (circle) : \citet{Gnerucci2011}, \citet{Epinat2012}, \citet{Livermore2015}, \citet{Wisnioski2015}, \citet{Contini2016}, \citet{Swinbank2017}, \citet{Turner2017}, \citet{Johnson2018}, and \citet{Girard2018a}. Only the velocity dispersion was available from \citet{Wisnioski2015}. The mean redshift values of each redshift bin in the right panels are 0.8 (orange), 1.4 (dark blue), and 3.3 (red) corresponding to redshift bins of $z<1$, $1<z<3$, and $z>3$, respectively. The circles represent the mean velocity dispersion (top right panel) and mean $\upsilon_{rot}/\sigma_0$ (bottom right panel) of each stellar mass bin : 8.0< log(\mstar/\msun)<8.7, 8.7< log(\mstar/\msun)<9.1, 9.1< log(\mstar/\msun)<9.8, 9.8< log(\mstar/\msun)<10.2,  and  log(\mstar/\msun)>10.2. The error bar on these circles is the error on the mean (with a confidence level of 98\%). The large black squares in the left panels represent the mean of the two mass bins of KLASS for galaxies with log(\mstar/\msun)<9.5 and log(\mstar/\msun)>9.5. The dashed line shows the best stellar mass and redshift dependent fit and on the KLASS sample (described in Sect. \ref{sect:62}). The grey solid lines are draws from the MCMC samples. We also present $\sigma_0$ and $\upsilon_{rot}/\sigma_0$ as a function of redshift in Fig. \ref{sigma_vs_redshift} in Appendix \ref{A2}.}
\label{sigma_vs_mass}
\end{figure*}

The rotation velocity as a function of stellar mass is plotted in Fig. \ref{TF_klass} (left). We present the galaxies that are rotation-dominated, with a regular rotation (in orange) and irregular rotation (in green), and the dispersion-dominated galaxies (in blue). For each sub-sample, we fit a linear curve with a fixed slope of  $m=0.274$ corresponding to the one obtained by \citet{Reyes2011} for galaxies in the local Universe:
\begin{equation}
\mathrm{log}(\upsilon) =b + m[ \mathrm{log}(\mstar/\msun)-10.10]
\end{equation}
where $b$ is the zero-point. \citet{Reyes2011} found a value of $ b= 2.127$ for galaxies in the local Universe with a Chabrier IMF. We find a velocity zero-point offset from the local relation of  $-0.40$ dex for the dispersion-dominated galaxies, $-0.02 $ dex for the irregular rotators, and $+0.22$ dex for the regular rotators. 
The irregular rotators show a value similar to the one of the local galaxies and the regular rotation-dominated galaxies have a positive zero-point offset compared to the local relation. 
The negative offset obtained for the dispersion-dominated galaxies likely indicates that the pressure support in the disk is significant for the galaxies showing a low $\upsilon_{rot}/\sigma_0$ ratio, and that the rotation velocity does not trace well the dynamical mass. To take into account for the pressure support, we use the circular velocity, $V_{circ}$, which combines both the rotation velocity and intrinsic velocity dispersion to trace the dynamical mass in galaxies:

\begin{equation}
V_{circ} = \sqrt{ \upsilon_{rot}^2 + 3.4\sigma_0^2},
\label{eq2}
\end{equation}
where the value 3.4 is obtained by assuming an exponential disk at a radius  equal to the effective radius, $R_e$ \citep[e.g.,][]{Burkert2010, Price2019}. We use this circular velocity, $V_{circ}$, to plot the Tully-Fisher in Fig. \ref{TF_klass} (right). In this plot, we use again the linear fit obtained by \citet{Reyes2011} to compare with our data. We assume that the pressure support is negligible compared to the velocity rotation for these local galaxies. Indeed, if we take a value of $\sigma_0\sim10-15$ \kms \, typical of these local galaxies \citep{Pizagno2007} in Eq. \ref{eq2}, we find that $V_{circ}\simeq \upsilon_{rot}$, except for galaxies with log(\mstar/\msun)$\lesssim8.0$.
We recover a tight correlation (black line), similar to \citet{Kassin2012}, \citet{Simons2016}, and \citet{Turner2018}. We find an offset of  $+0.18$ dex for the full sample which is consistent with the value we obtain for the regular rotators when no correction is applied for the pressure support (orange line in Fig. \ref{TF_klass} (left)) and the value obtained by \citet{Turner2018} for galaxies at $z\sim3.5$. A positive offset can be interpreted as galaxies having a higher gas fraction at this epoch \citep[e.g.,][]{ Tacconi2013} since they still have to convert most of their gas into stars. 

Using the circular velocity, we find a scatter of 0.2 dex. This scatter is defined as the standard deviation of the residuals on the Tully-Fisher relation. \citet{Reyes2011} obtained a value of 0.056 dex for the local relation. Similarly to other high-redshift surveys, the scatter around the relation is much higher than in the local Universe. This could be due to the larger uncertainties on the rotation velocity and velocity dispersion that we obtain at high redshift. For our sample, we find that the observational uncertainties contribute to $\sim0.06$ dex, which is lower than the observed scatter around the relation.
It could indicate that galaxies are at different evolutionary stages at this epoch, with some of them that already have converted most of their gas into stars while others only started to build their disk and form stars. 

These data do not suggest that the slope has any dependence on the stellar mass and redshift when using the circular velocity, which is in good agreement with the results from \citet{Harrison2017}, for example, but in contradiction with \citet{Christensen2017}, where they reported a steeper slope for low-mass galaxies. However, this sample is small and more statistics are needed to better look at the variation of the slope with redshift and stellar mass.

\subsection{Evolution of the kinematic properties}
\label{sect:62}
The relation between the kinematic properties and stellar mass has recently been analyzed by several surveys \citep[e.g.,][]{Kassin2012,Simons2017,Mason2017,Girard2018a}. A correlation between $\sigma_0$ and stellar mass was reported by some studies \citep[e.g.,][]{Johnson2018}, while others did not find any correlation \citep[e.g.,][]{Simons2017, Turner2017}. Similarly, a correlation between $\upsilon_{rot}/\sigma_0$ and stellar mass was found by several surveys \citep[e.g.,][]{Kassin2012,Simons2017}, but others reported no clear trend \cite[e.g.,][]{Contini2016}.
Previous surveys were also limited to a small number of low-mass galaxies with only a few with log(\mstar/\msun)<10 at $z>1$ (see Figs. \ref{sigma_vs_mass} and \ref{sigma_vs_redshift}). In this section, we use KLASS to investigate the trend between the stellar mass and $\sigma_0$ and $\upsilon_{rot}/\sigma_0$. Due to the limited number of galaxies in our sample, we also combine our sample with galaxies publicly available from previous surveys.

The velocity dispersions for the full KLASS sample are similar to other values reported in the literature at similar redshift (see the top panels of Figs. \ref{sigma_vs_mass} and \ref{sigma_vs_redshift}). We obtain mean and median values of 50 \kms \, and 40 \kms \, for this sample, respectively. When dividing the sample in two mass bins, we find that the galaxies with a stellar mass of log(\mstar/\msun)<9.5 (and median mass of  log(\mstar/\msun)=9.1) have a mean and median velocity dispersion of 35 \kms \, and 29 \kms. We obtain for the more massive galaxies with log(\mstar/\msun)>9.5 (and median mass of log(\mstar/\msun)=9.9) a mean and median velocity dispersion of 64 \kms \, and 55 \kms (see Fig. \ref{sigma_vs_mass}). We fit a model of the form $\sigma_0\sim(\mstar/10^{10}M_\odot)^\alpha \times (1+z)^\beta$ using Bayesian inference, sampling the posterior with EMCEE \citep{Foreman-Mackey2013} to characterize the redshift and mass evolution. We use a flat prior on $\alpha$ between -2 and 2, $\beta$ between -1 and 1, and a flat prior on the normalization factor $10^\gamma$ where $0 \leq \gamma \leq 4$. We also fit for a constant rescaling of the errors on the velocity dispersion, to account for systematic underestimation of the errors, and marginalise over this. We find that the velocity dispersion is strongly correlated to the stellar mass with $\alpha=0.22\pm0.05$, with no significant evolution with redshift: $\beta=0.12\pm0.43$. The best fit is shown as a dashed line in Fig. \ref{sigma_vs_mass} (top left panel) for the mean redshift of our sample ($z\sim1.3$; see also Fig. \ref{sigma_vs_redshift} for the relation as a function of redshift).

To further explore this trend, we combine our data from KLASS with several surveys from the literature: \citet{Gnerucci2011}, \citet{Epinat2012}, \citet{Livermore2015}, \citet{Wisnioski2015}, \citet{Contini2016}, \citet{Swinbank2017}, \citet{Turner2017}, \citet{Johnson2018} and \citet{Girard2018a}. This allows us to get enough statistics to study in details the trend between the velocity dispersion, $\upsilon_{rot}/\sigma_0$, and the stellar mass and redshift. We note that several of these studies have a different spatial and spectral resolution and used different sample selection and methods to measure the kinematic properties. This could potentially be a source of scatter and bias the interpretation. For example, the MASSIV survey \citep{Epinat2012} used an error-weighted mean of the beam-smearing corrected velocity dispersion map. This method is known to give slightly higher velocity dispersions than other methods \citep{Davies2011}. However, most of the surveys that we combine used kinematic modeling to determine the velocity dispersion or measured this property in the outskirt of galaxies. These two methods give similar results (see Sect. \ref{sec:veldispersion} and Fig. \ref{sigma_comparison_klass}). Moreover, most of the galaxies in the compilation have been observed with KMOS, and therefore have a similar spatial and spectral resolution \citep{Wisnioski2015, Swinbank2017, Turner2017, Johnson2018, Girard2018a}. We also note that most of the galaxies in this compilation are at $z\sim1$ (see Fig. \ref{sigma_vs_redshift}) due to the large number of galaxies in the KROSS sample \citep{Johnson2018}.

Fig. \ref{sigma_vs_mass} (top right panel) shows the dispersion as a function of stellar mass for the galaxies included in all the surveys listed above. To investigate the dependence on redshift, we divide the sample in the following redshift bins:  $z<1$, $1<z<3$, and $z>3$. The mean values of each bin is 0.8, 1.4, and 3.3, respectively. We consider four stellar mass bins: 8.7 < log(\mstar/\msun)< 9.1, 9.1 < log(\mstar/\msun)< 9.8, 9.8 <log(\mstar/\msun) < 10.2, and log(\mstar/\msun) > 10.2. The galaxies from KLASS allow us to add statistics in the two lower mass bins at $z\sim1-2$. We therefore focus on the mass evolution at log(\mstar/\msun)<10 in this work. For galaxies at $z\sim0.8$ the velocity dispersion increases from $\sim35$ to $\sim45$ \kms \, for galaxies at log(\mstar/\msun)$\sim8.9$ to 10.5. This is consistent with a positive correlation between these quantities. There is also an increase of the velocity dispersion with stellar mass at $z>1$, but the correlation is less obvious at these redshifts due to the large error bars. This is consistent with the results obtained by \citet{Johnson2018} where they combined data from the KROSS sample at $z\sim0.9$, the SAMI survey at $z\sim0.05$ and a sample from MUSE at $z\sim0.5$. This correlation could be partly due to an effect of residual beam-smearing since it affects more the massive galaxies \citep{Burkert2016, Johnson2018}. However, \citet{Pillepich2019} found a similar trend using IllustrisTNG simulations. They obtained that the H$\alpha$ velocity dispersion of galaxies increases with stellar mass, for all redshifts. This is also consistent with the $\sigma_0 - \mstar$ relation, derived in \citet{Girard2018a} assuming a quasi-stable disk (Toomre parameter $Q=1$) and using the Tully-Fisher relation, which predicts an increase of the intrinsic velocity dispersion with stellar mass.

We also find an increase of the dispersion with redshift, as found by many other surveys \citep[e.g.,][]{ForsterSchreiber2009, Wisnioski2015}. It is however still unclear if an increase with redshift is seen for the lower mass galaxies (log(\mstar/\msun)$\lesssim $9.1) as the values are similar for $0.5<z<1.0$ and $1.0<z<3.0$ (see Fig. \ref{sigma_vs_mass}). This might also only be due to the lack of observations of low-mass galaxies at higher redshift ($z>1$). 

We also plot $\upsilon_{rot}/\sigma_0$ as a function of stellar mass in Fig. \ref{sigma_vs_mass} (bottom panels). Using the same fitting method as in Fig. \ref{sigma_vs_mass} (left panels) for our galaxies from KLASS, we find no significant evolution of $\upsilon_{rot}/\sigma_0$ with stellar mass ($\alpha=0.03\pm0.07$) or redshift ($\beta=-0.26\pm0.45$). The best fit is shown for the mean redshift of our sample ($z\sim1.3$).
Using the same compilation from the literature and same stellar mass and redshift bins as previously described in Fig. \ref{sigma_vs_mass} (top right panel), we see a correlation between $\upsilon_{rot}/\sigma_0$ and stellar mass with an increase of this ratio for the more massive galaxies at $0.5<z<1.0$.  However, this correlation is not clearly seen at $z>1$.
An increase of $\upsilon_{rot}/\sigma_0$ with stellar mass means that low-mass galaxies are less rotation-dominated at this redshift. This suggests that more massive galaxies settle first into regular rotating disks \citep[e.g.][]{Kassin2012,Simons2016,Johnson2018}. 
Similarly, \citet{Pillepich2019} found a correlation of $\upsilon_{rot}/\sigma_0$ with stellar mass at $z<1$ from their simulations. The trend between $\upsilon_{rot}/\sigma_0$ and stellar mass flattens at $z>1$ according to their study as it might be seen in Fig. \ref{sigma_vs_mass}. 
We also find that $\upsilon_{rot}/\sigma_0$  decreases with redshift for all masses, which agrees with the results from other observations \citep[e.g.][]{Simons2017, Wisnioski2015} and simulations \citep{Pillepich2019}. This indicates that galaxies are more turbulent and pressure supported at high redshift.

\begin{figure}
\centering
\includegraphics[width=\columnwidth]{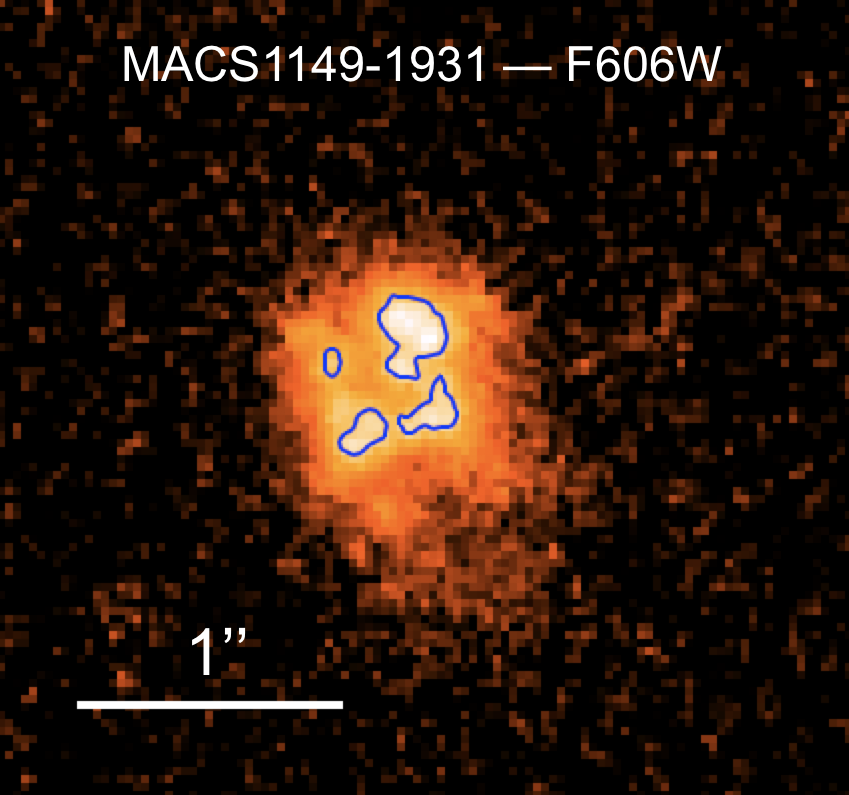}
%
%
\caption[]{F606W \textit{HST} image of the galaxy MACS1149-1931. The blue contours show the clumps detected with Astrodendro.}
\label{example_clumps_detection}
\end{figure}

\begin{figure*}
\centering
\includegraphics[scale=0.68]{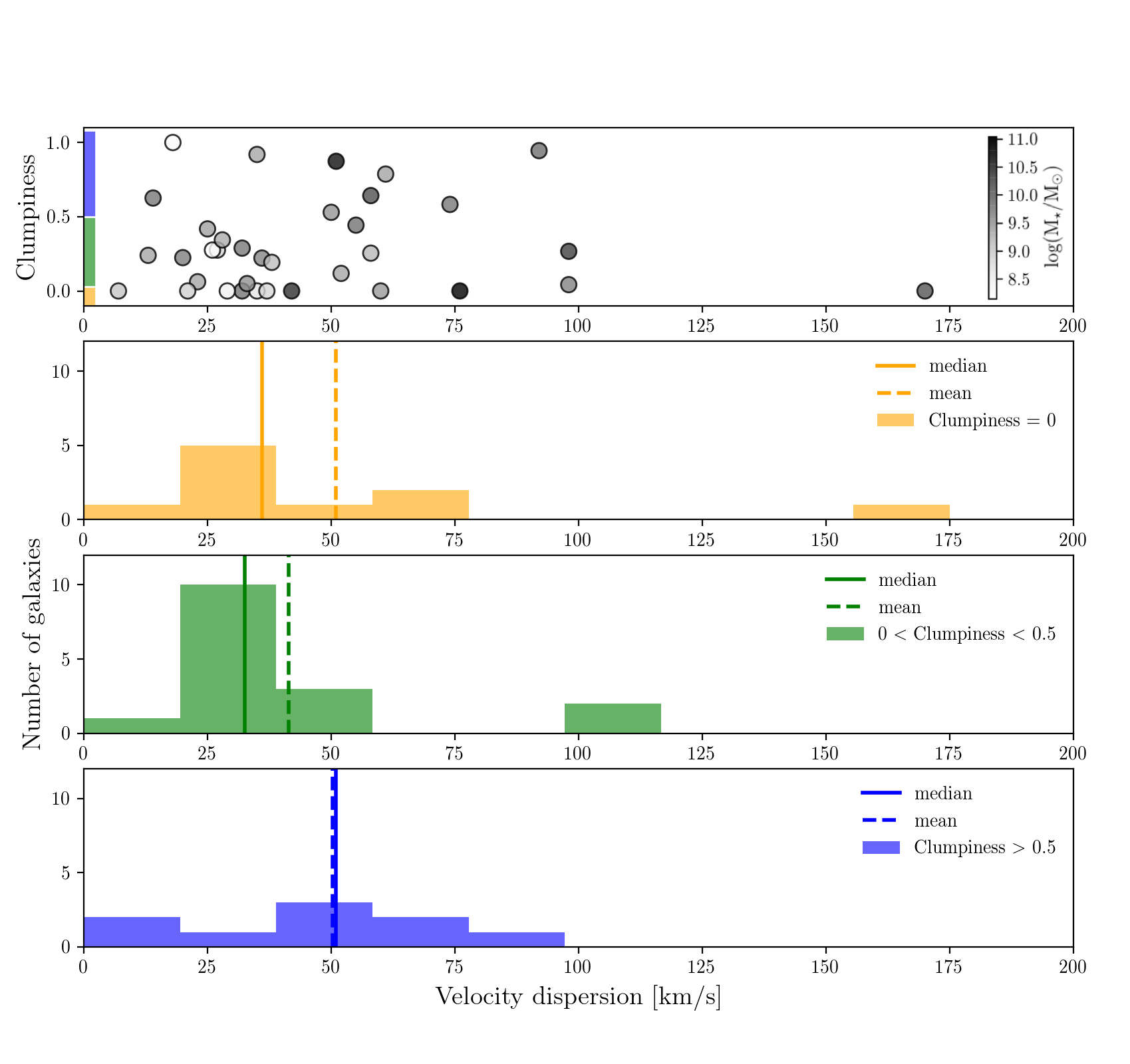}
%
%
\caption[]{Clumpiness as a function of the velocity dispersion (top panel) and histograms of the velocity dispersion for the galaxies with no clumps (middle top panel), a clumpiness between 0 and 0.5 (middle bottom panel), and a clumpiness higher than 0.5 (bottom panel). The dashed and solid lines represent the mean and median of each distribution, respectively. The top panel is color-coded to indicate the corresponding stellar mass and the colors on the left correspond to the clumpiness bins of each histogram. }
\label{clump1}
\end{figure*}
\begin{figure*}
\centering
\includegraphics[scale=0.68]{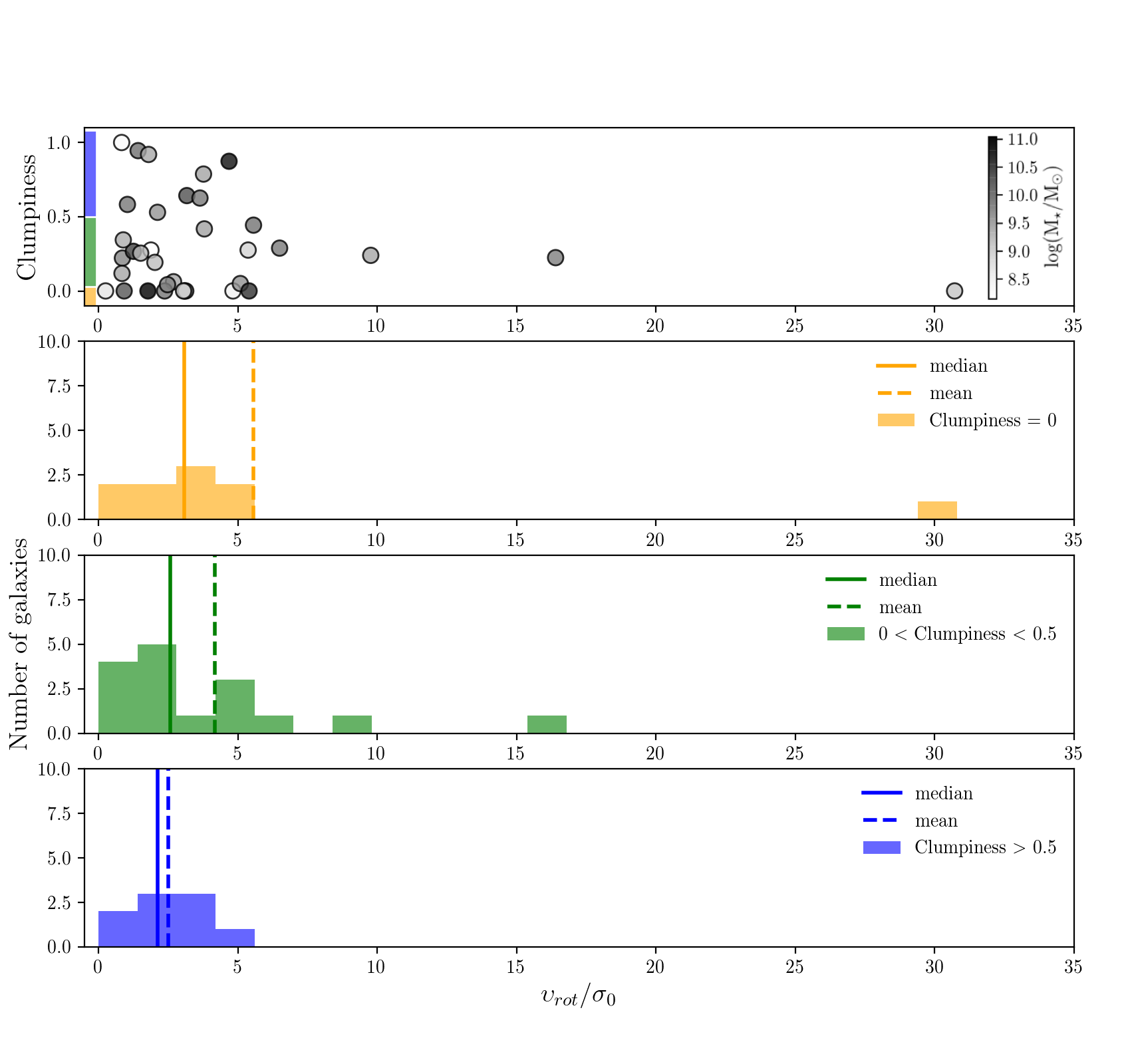}
%
%
\caption[]{Clumpiness as a function of $\upsilon_{rot}/\sigma_0$ (top panel) and histograms of $\upsilon_{rot}/\sigma_0$ for the galaxies with no clumps (middle top panel), a clumpiness between 0 and 0.5 (middle bottom panel), and a clumpiness higher than 0.5 (bottom panel). The dashed and solid lines represent the mean and median of each distribution, respectively. The top panel is color-coded to indicate the corresponding stellar mass and the colors on the left correspond to the clumpiness bins of each histogram. }
\label{clump2}
\end{figure*}

\subsection{Clumpiness and kinematics properties}

It is now well established that galaxies at cosmic noon have clumps that are forming stars at rates 100$\times$ higher than in local star forming regions of similar size \citep[e.g.,][]{Genzel2011,Guo2012}. In this section, we explore the physical conditions in which these clumps possibly formed in the disks by comparing the rest-frame UV clumpiness to the kinematics properties of the galaxies.

To determine the clumpiness of the galaxies we use the F435W, F606W, and F775W \textit{HST} images tracing the B, V,  and i bands which correspond to the rest-frame UV images of galaxies at redshift $0.5<z<1$, $1<z<2$, $2<z<3$, respectively \citep{Guo2012}.
We identify the clumps in every galaxy of the sample using Astrodendro\footnote{\url{http://www.dendrograms.org}}. This tool uses dendrograms, which are trees that represent hierarchically the different structures in the data  \citep{Rosolowsky2008}. The code finds the brightest value in the image and associates the spaxels around it to this clump until a new local maximum is found. There are two input parameters: the minimum flux for which the spaxels are considered and the minimum value between the maxima of two independant clumps. We fix the minimum value at 3$\sigma$ above the noise level and the minimum value between the peaks of two clumps at 1$\sigma$ \citep{Rosolowsky2008}. We only consider clumps that have at least a size comparable to the PSF and that are off-center. An example of the clumps detected with Astrodendro in the galaxy MACS1149-1931 is shown in Fig. \ref{example_clumps_detection}.

In this work, we define the clumpiness as the ratio between the UV light in all the clumps and the total UV light of the galaxy, $clumpiness = \Sigma \, L_{clump}/L_{total} $ \citep[e.g.][]{Soto2017, Messa2019}. We sum the UV light of clumps detected with Astrodendro and divide by the total UV flux. 
We find that 23\% of the galaxies in our sample are not clumpy (no clump detected). We detect clumps in 58\% of the galaxies, with 21\% and 37\% that have a clumpiness higher than 0.5 and between 0 and 0.5, respectively. The rest of the galaxies are potential mergers for which it is not clear from their \textit{HST} images and kinematic maps if the clumps detected are actual star-forming clumps or distinct objects in interactions.  The clumpiness fraction of $\sim58\%$ found for this sample is in agreement with \citet{Shibuya2016}, where they obtained a fraction between $50\%$ and $70\%$ for galaxies at $0.5<z<3$. The clumpiness parameter is listed in Table \ref{kin_properties}.

Fig. \ref{clump1} presents the clumpiness as a function of the velocity dispersion, color-coded with their stellar mass, with the histograms of the velocity dispersion for the galaxies showing no clump, the galaxies having a clumpiness between 0 and 0.5, and the galaxies with a clumpiness higher than 0.5. The clumpy galaxies show a wide range of dispersion, typically between 15 and 100 \kms.  We obtain a Spearman rank correlation coefficient of 0.04 and a probability of an uncorrelated system (p-value) of 84\%, which indicates that there is no trend between the clumpiness and velocity dispersion.  Also, we do not see a strong dependence between the clumpiness and the stellar mass in this sample (see top panel of Figs. \ref{clump1} and Fig. \ref{clump2}), therefore the dependence in stellar mass should not affect our results. However, the galaxies with a clumpiness higher than 0.5 show a median velocity dispersion of $51\pm 18$ \kms, which is higher than the median of $36\pm 13 $ \kms \, and $33\pm 14$ \kms \, obtained for the galaxies with no clump and a low clumpiness between 0 and 0.5, respectively.

Fig. \ref{clump2} presents the clumpiness as a function of $\upsilon_{rot}/\sigma_0$, with the histograms of  $\upsilon_{rot}/\sigma_0$ for the galaxies with no clump, the galaxies with a clumpiness between 0 and 0.5, and the galaxies with a clumpiness higher than 0.5. We find a  Spearman rank correlation coefficient of -0.06 and a p-value of 74\%, which indicates that there is no clear trend between the clumpiness and $\upsilon_{rot}/\sigma_0$. The three clumpiness categories all show a median of  $\upsilon_{rot}/\sigma_0\sim 2-3$.

\citet{Fisher2017} found an anti-correlation between the clumpiness and $\upsilon_{rot}/\sigma_0$ for their massive galaxies at $z\sim0.1$ analog to high-redshift galaxies. \citet{Messa2019} obtained a weak trend between the clumpiness and $\upsilon_{rot}/\sigma_0$ for their UV-selected star-forming galaxies in the Lyman-Alpha Reference Sample (LARS) at $0.03<z<0.2$, but with a large scatter. In this work, we do not find any clear trend, but we also look at very different types of galaxies, with lower stellar mass and at higher redshift.
The higher velocity dispersion observed for the most clumpy galaxies is however consistent with clumps that would form by fragmentation in a turbulent, gas-rich disk due to gravitational instabilities \citep{Tamburello2015}. The large scatter and clumpy galaxies with higher $\upsilon_{rot}/\sigma_0$ seen in this sample could be due to turbulent, gas-rich galaxies that stabilize after forming their clumps and evolve into less turbulent and more stable disk galaxies with clumps that are still present in their internal structure. This could also indicate that clumps can form in very different physical conditions, which would be  inconsistent with clumps that form by gravitational instabilities \citep[e.g.,][]{Meng2019}.

Additionally, the low $\upsilon_{rot}/\sigma_0$ and relatively high velocity dispersion observed in some galaxies that do not have any clumps suggest that either the turbulence originates from other processes in these particular objects or that these galaxies could be about to form clumps.

\section{Conclusion}
\label{sec:conclusion}

We have presented results from the full KLASS sample, a survey that uses gravitational lensing to study the resolved kinematics in 44 star-forming galaxies with a median stellar mass of log(\mstar/\msun)$\sim9.5$ at $0.6<z<2.3$. The main results of this work are :

\begin{enumerate}
   
   \item From the 44 galaxies with resolved kinematics, we find that $39\pm9$\% show regular rotation with $\upsilon_{rot}/\sigma_0$>3, $27\pm7$\% are irregular rotators with $1<\upsilon_{rot}/\sigma_0<3$, $16\pm6$\% are dominated by dispersion with $\upsilon_{rot}/\sigma_0$<1 or do not show any velocity gradient, and $18\pm7$\% are potential mergers. This means that most of the galaxies show rotation, but are barely dominated by the rotation.
   
   \item We find that the KLASS sample follows the Tully-Fisher relation, after adding the contribution for the pressure support in the galaxies (traced by the velocity dispersion). We obtain a positive zero-point offset of +0.18 dex and a larger scatter of 0.2 dex when comparing to the relation from the local Universe \citep{Reyes2011}. This could suggest that these galaxies are more gas-rich and are at different evolutionary stages at this epoch, with some galaxies only starting to convert their gas into stars while other are more advanced in the process and have already converted a larger fraction of their gas.
   
   \item We find a strong evolution of the velocity dispersion with stellar mass in the KLASS sample, with a median value of 29 \kms \, and 55 \kms \, for the galaxies with log(\mstar/\msun)<9.5 and log(\mstar/\msun)>9.5, respectively. 
   When combining our data with other studies from the literature, we still find an increase of the velocity dispersion with stellar mass and an increase of the dispersion with redshift. This evolution of the velocity dispersion with redshift for the lower mass galaxies (log(\mstar/\msun)<9.1) is however still unclear, possibly due to a lack of observations of these galaxies at higher redshift. 
   We also find an increase of $\upsilon_{rot}/\sigma_0$ with stellar mass for galaxies at $0.5<z<1.0$, and a weak increase or flat trend for $z>1$. This could indicate that low-mass galaxies settle into a more regular rotating disk after the more massive galaxies at $0.5<z<1$. 
   
    \item We find that the most clumpy galaxies in our sample have a slightly higher velocity dispersion on average, but we do not find any clear trend between the UV clumpiness of the galaxies and the velocity dispersion and  $\upsilon_{rot}/\sigma_0$. The diversity of the kinematics properties found for the clumpy galaxies could be due to turbulent, gas-rich galaxies that evolve into less turbulent and more stable disk after forming clumps in their disk. This could also indicate that clumps can form in different physical conditions. The galaxies with no clump also show a wide range of dispersion ($10<\sigma_0 [$\kms$] <75$). Several of them also have a low $\upsilon_{rot}/\sigma_0$ ratio ($<3$). This could mean that the turbulence in these galaxies is either due to different mechanisms than in the clumpy galaxies or that these galaxies are about to form clumps.

\end{enumerate}


\section*{Acknowledgements}

MG and DBF acknowledge support from Australian Research Council (ARC) DP grant DP160102235 and Future Fellowship FT170100376.
CAM acknowledges support by NASA Headquarters through the NASA Hubble Fellowship grant HST-HF2-51413.001-A awarded by the Space Telescope Science Institute, which is operated by the Association of Universities for Research in Astronomy, Inc., for NASA, under contract NAS5-26555.
GLASS was supported by NASA through STSCI grant HST-GO-13459.
This research made use of astrodendro, a Python package to compute dendrograms of Astronomical data (http://www.dendrograms.org/).




\bibliographystyle{mnras}
\bibliography{mgirard_KLASS2020.bbl} 




\appendix
\section{Kinematic maps}
\label{A1}

Figure \ref{map_galaxies} shows the HST images, H$\alpha$, [OIII] or [OII] flux maps, the velocity dispersion maps, the observed velocity maps, the velocity maps from the models, the rotation curves and dispersion profiles of each galaxy showing a velocity gradient in the KLASS sample. 

\begin{figure*}
\includegraphics[scale=0.82]{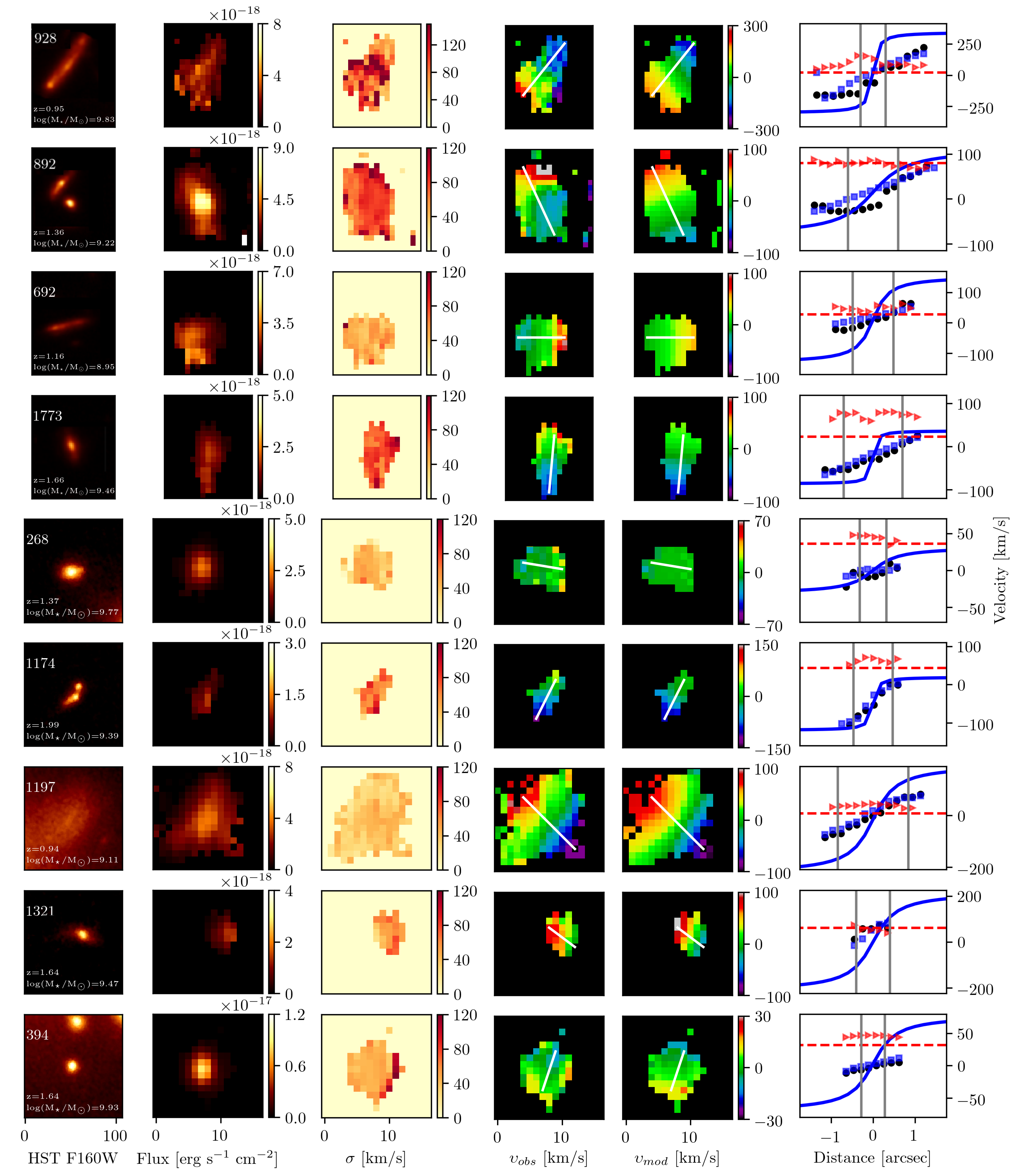}

%
%
\caption[]{HST/WFC3 F160W near-infrared images, H$\alpha$, [OIII] or [OII] flux in \ergscm maps, velocity dispersion maps, observed velocity maps, and velocity maps from the kinematic models of 42 galaxies from KLASS. The white lines in the velocity maps indicate the major axis. In the right panels, the black circles and blue squares represent the rotation curves extracted on the major axis of the observed velocity maps and velocity maps from the kinematic models, respectively. The blue solid lines show the intrinsic rotation curves from the models (corrected for the inclination). The vertical grey lines indicate the value of the effective radius obtained with GalPaK$^{\rm3D}$. The red triangles and red dashed lines represent the velocity dispersion profiles and the intrinsic velocity dispersion obtained from the kinematic models, respectively.}
\label{map_galaxies}
\end{figure*}

\begin{figure*} 
\centering
\setcounter{figure}{0}
\includegraphics[scale=0.86]{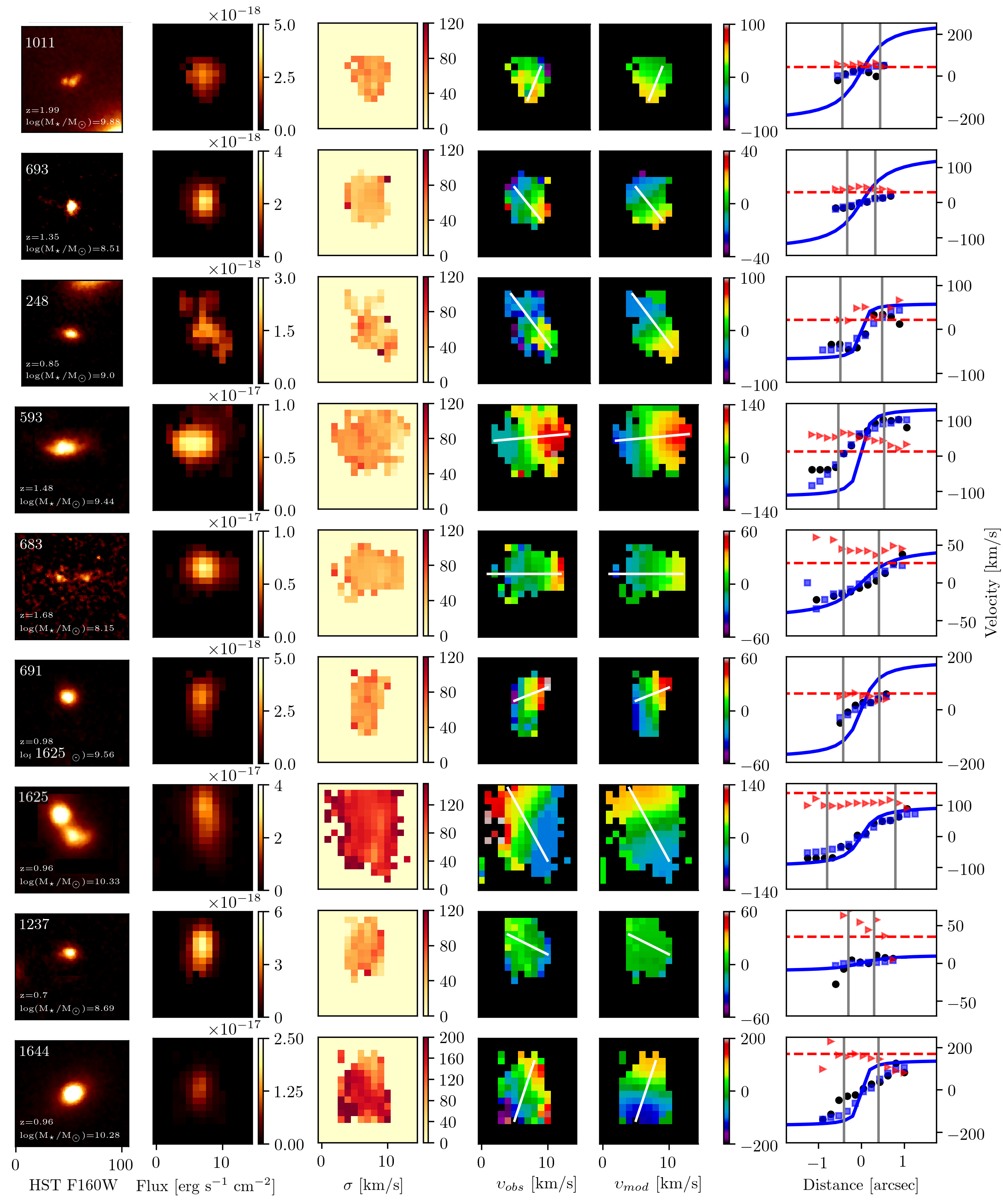}
\caption{Continued.}
\end{figure*}

\begin{figure*} 
\centering
\setcounter{figure}{0}
\includegraphics[scale=0.82]{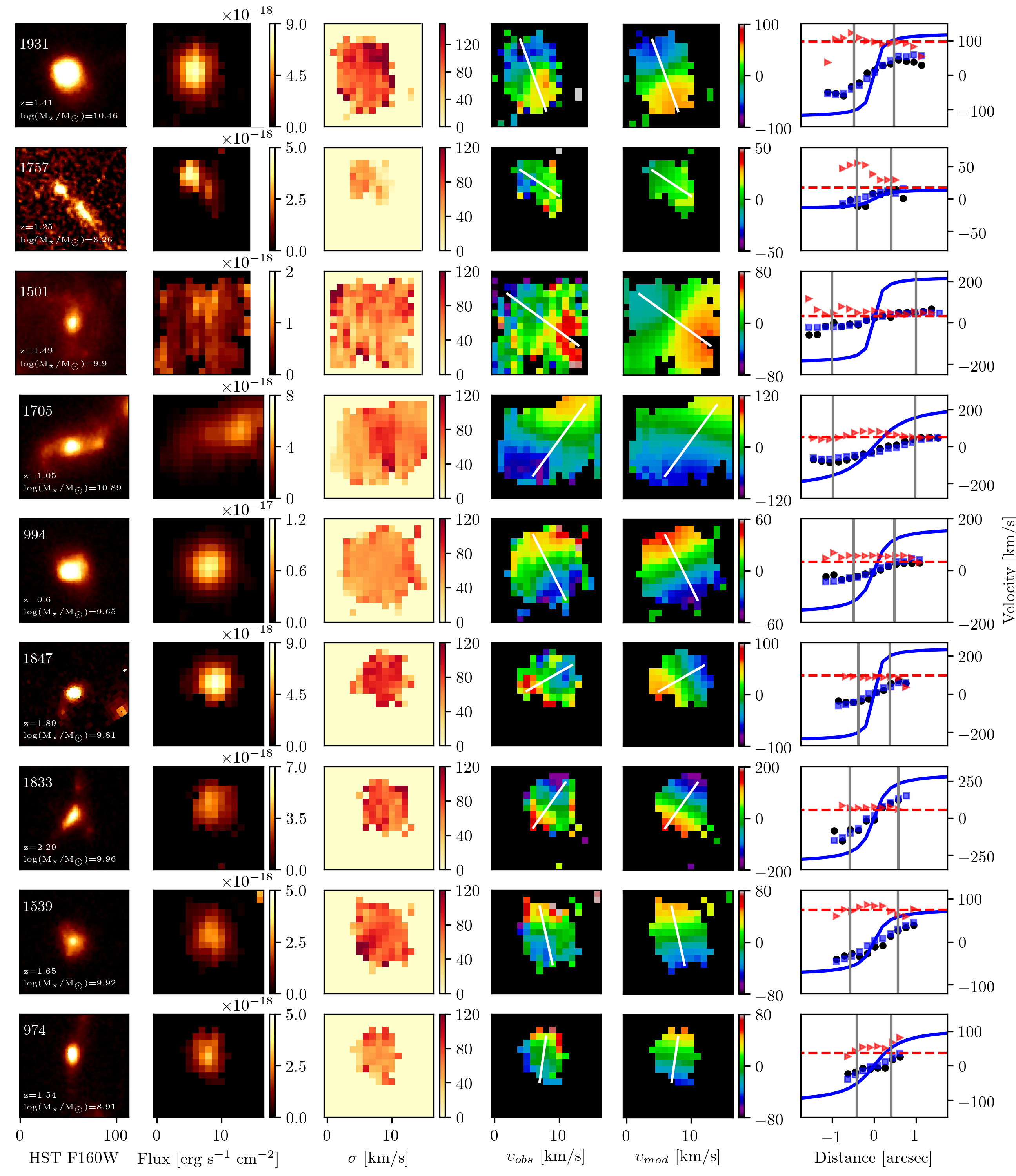}
\caption{Continued.}
\end{figure*}

\begin{figure*} 
\centering
\setcounter{figure}{0}
\includegraphics[scale=0.81]{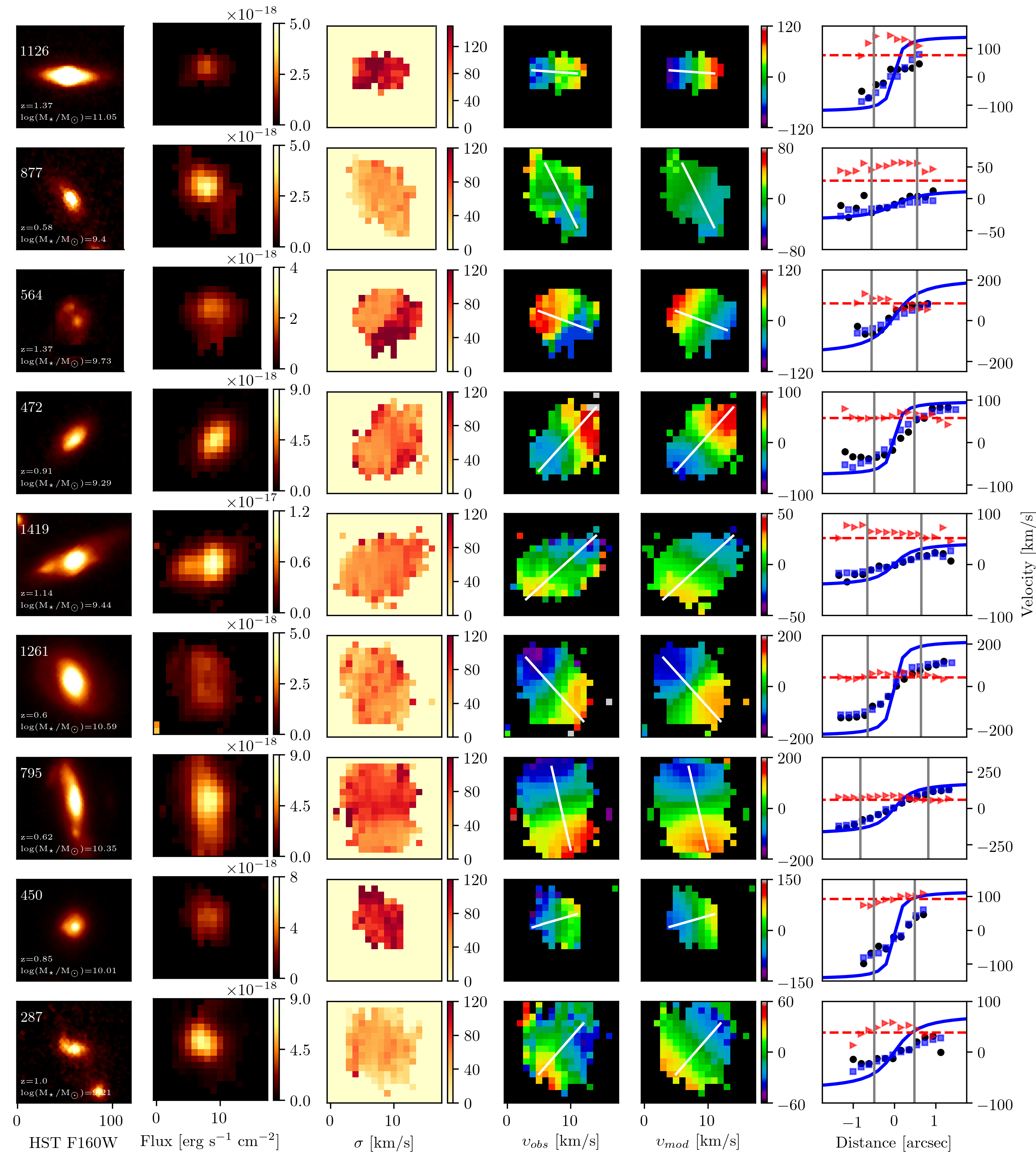}
\caption{Continued.}
\end{figure*}

\begin{figure*} 
\centering
\setcounter{figure}{0}
\includegraphics[scale=0.82]{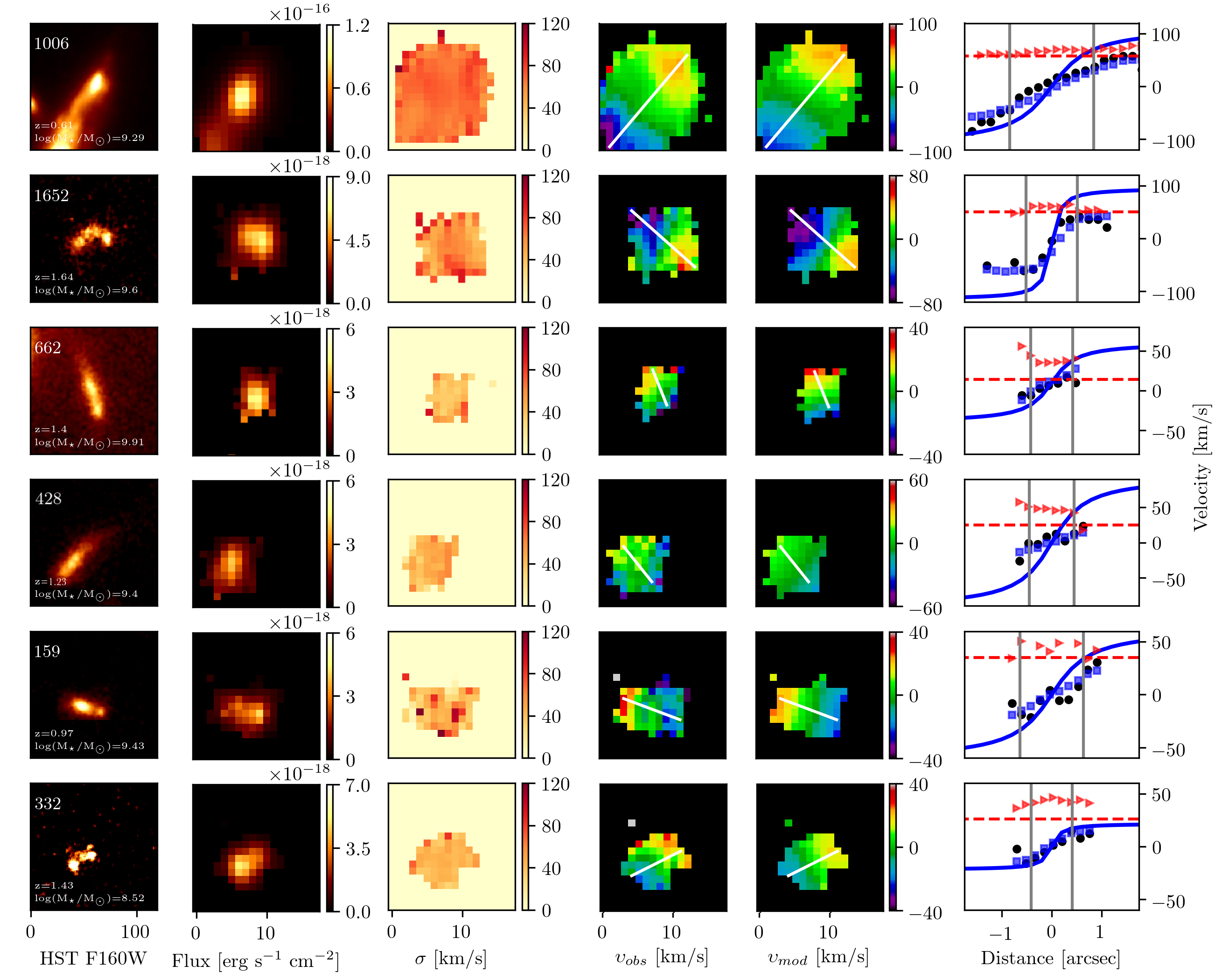}
\caption{Continued.}
\end{figure*}

\section{Kinematic properties as a function of redshift}
\label{A2}
\begin{figure*}
\includegraphics[scale=0.9]{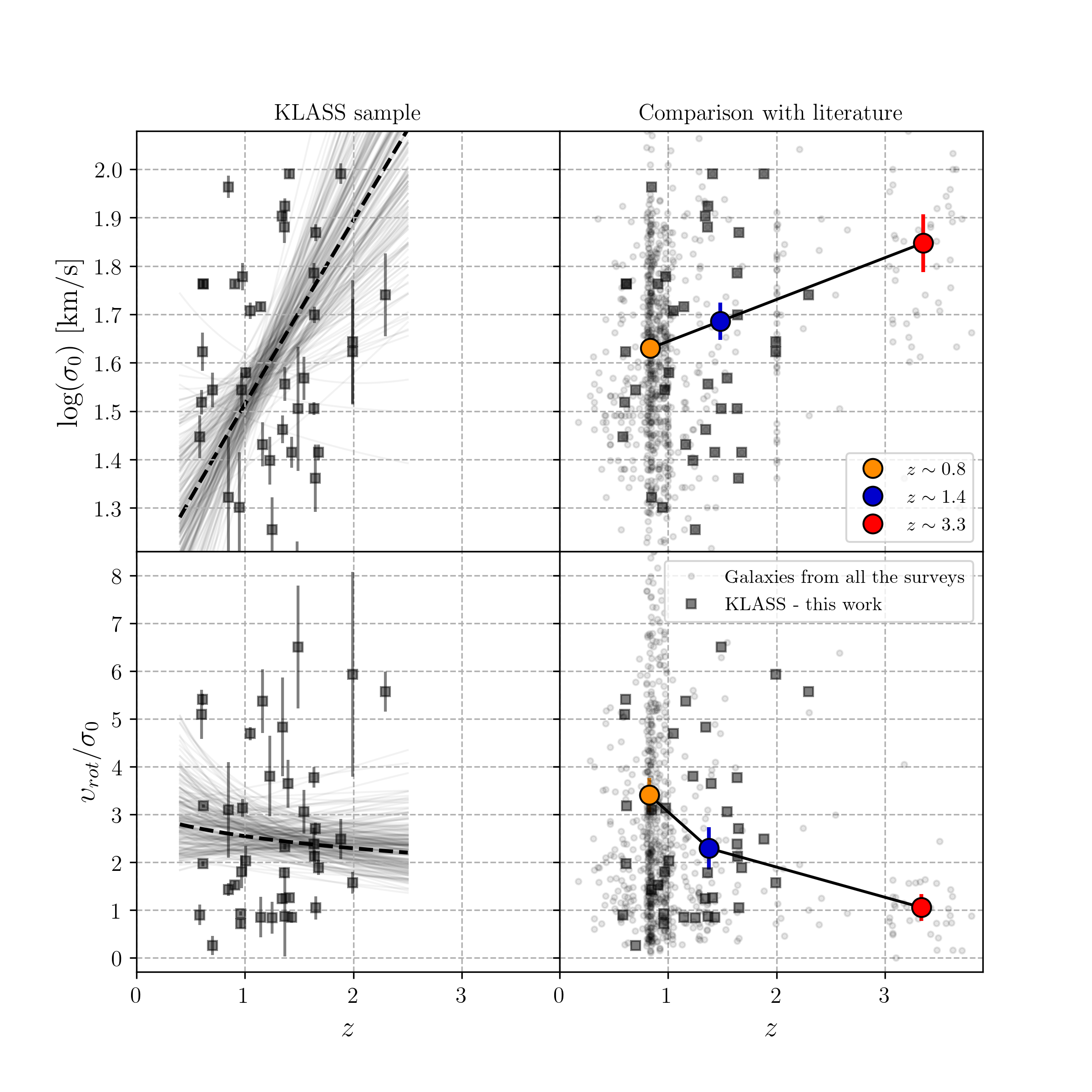}
\medskip
%
\caption[]{Intrinsic velocity dispersion and rotation velocity to intrinsic velocity dispersion ratio, $\upsilon_{rot}/\sigma_0$, as a function of redshift for KLASS (square) and a compilation of galaxies from different surveys (circle) : \citet{Gnerucci2011}, \citet{Epinat2012}, \citet{Livermore2015}, \citet{Wisnioski2015}, \citet{Contini2016}, \citet{Swinbank2017}, \citet{Turner2017}, \citet{Johnson2018}, and \citet{Girard2018a}. Only the velocity dispersion was available from \citet{Wisnioski2015}. The mean redshift values of each redshift bin are 0.8 (orange), 1.4 (dark blue), and 3.3 (red) corresponding to redshift bins of $z<1$, $1<z<3$, and $z>3$, respectively. The colors for the redshift bins are the same as in Fig. \ref{sigma_vs_mass}. The circles represent the mean velocity dispersion (top right panel) and mean $\upsilon_{rot}/\sigma_0$ (bottom right panel) of each redshift bin. The error bar on these circles is the error on the mean (with a confidence level of 98\%). The dashed line on the left panels shows the best stellar mass and redshift dependent fit on the KLASS sample (described in Sect. \ref{sect:62}).The grey solid lines are draws from the MCMC samples.}
\label{sigma_vs_redshift}
\end{figure*}

Figure \ref{sigma_vs_redshift} shows the intrinsic velocity dispersion and the rotation velocity to intrinsic velocity dispersion ratio, $\upsilon_{rot}/\sigma_0$, as a function of redshift.

\bsp	
\label{lastpage}
\end{document}